\documentclass{article}
\usepackage[utf8]{inputenc}
 \usepackage{bm}
\usepackage{amsmath}
\usepackage{lscape}
\usepackage{multirow}
\usepackage[margin=2cm]{caption}
\usepackage[round]{natbib}
\usepackage{authblk}
\usepackage{listings}
\usepackage{array}
\usepackage{soul}
\usepackage[usenames,dvipsnames]{color}   
\usepackage{float}
\usepackage{colortbl}
\usepackage{amssymb}
\usepackage[figuresright]{rotating}
\usepackage{url}
\usepackage[dvipsnames]{xcolor}
\usepackage{graphicx} %For loading graphic files
\usepackage{titlesec}
\usepackage[left=1.5cm, right=1.5cm, top=2cm, bottom=2cm]{geometry}
\usepackage{changepage}
\usepackage{bbold}

\providecommand{\keywords}[1]
{
  \small	
  \textbf{\textit{Keywords: }} #1
}

\title{\textbf{Fast and flexible inference for joint models of multivariate longitudinal and survival data using Integrated Nested Laplace Approximations}}

\author{\textbf{Denis Rustand$^1$, Janet van Niekerk$^1$, Elias Teixeira Krainski$^1$,}\\ \textbf{H\aa vard Rue$^1$ and C\'ecile Proust-Lima$^2$}\\
\small \vspace{0.5cm} $^1$ Statistics Program, Computer, Electrical and Mathematical Sciences and Engineering Division, \hfill\\
King Abdullah University of Science and Technology (KAUST),\\
Thuwal 23955-6900, Kingdom of Saudi Arabia\\
$^2$ Univ. Bordeaux, Inserm, Bordeaux Population Health Center, UMR1219, F-33000 Bordeaux, France}

\date{}

\begin{document}

\maketitle

\abstract{\begin{adjustwidth}{40pt}{40pt}
\normalsize Modeling longitudinal and survival data jointly offers many advantages such as addressing measurement error and missing data in the longitudinal processes, understanding and quantifying the association between the longitudinal markers and the survival events and predicting the risk of events based on the longitudinal markers. A joint model involves multiple submodels (one for each longitudinal/survival outcome) usually linked together through correlated or shared random effects. Their estimation is computationally expensive (particularly due to a multidimensional integration of the likelihood over the random effects distribution) so that inference methods become rapidly intractable, and restricts applications of joint models to a small number of longitudinal markers and/or random effects. We introduce a Bayesian approximation based on the Integrated Nested Laplace Approximation algorithm implemented in the R package \textbf{R-INLA} to alleviate the computational burden and allow the estimation of multivariate joint models with fewer restrictions. Our simulation studies show that \textbf{R-INLA} substantially reduces the computation time and the variability of the parameter estimates compared to alternative estimation strategies. We further apply the methodology to analyze 5 longitudinal markers (3 continuous, 1 count, 1 binary, and 16 random effects) and competing risks of death and transplantation in a clinical trial on primary biliary cholangitis. \textbf{R-INLA} provides a fast and reliable inference technique for applying joint models to the complex multivariate data encountered in health research.\\\end{adjustwidth}}

\begin{adjustwidth}{40pt}{40pt} 
{Bayesian inference; Competing risks; Computational approaches comparison; Efficient estimation; Joint modeling; Multivariate longitudinal markers.}\end{adjustwidth}
\vspace{0.5cm}
\normalsize 
\section{Introduction}
It is common to observe longitudinal markers censored by a terminal event in health research with interest in the analysis of the longitudinal marker trajectory and the risk of the terminal event as well as their relationship. This includes the analysis of CD4 lymphocytes counts and AIDS survival \citep{wulfsohn1997joint}, prostate-specific antigen dynamics and the risk of cancer recurrence \citep{proust2009development}, cancer tumor dynamics and the risk of death \citep{Rustand20}, dynamics of aortic gradient and aortic regurgitations and their relationship with the competing risks of death or reoperation in the field of cardiac surgery \citep{Andrinopoulou14} or cognitive markers' relationship with the time to onset of Alzheimer's disease \citep{Kang22}, to name but a few.\\

Longitudinal markers are usually endogenous (i.e., their values are affected by a change in the risk of occurrence of the event) and measured with error, which prevents their inclusion as standard time-varying covariates in a survival model \citep{prentice1982covariate}. In this context, two-stage models have been introduced; first proposals were fitting a longitudinal model and subsequently using the estimated trajectory of the biomarker as a covariate in a survival model (\cite{self1992modeling, ye2008semiparametric}). While they accounted for measurement error of the biomarker in the survival model, they were failing to account for informative drop-out due to the survival event in the longitudinal submodel (\cite{albert2010estimating}). Two-stage models were then enhanced by fitting a longitudinal model at each event time to account for drop-out (\cite{de1994modelling, tsiatis1995modeling}). Beyond the heavy computational burden involved, the assumption of Gaussian distributed random effects at each event time was unreasonable (if the biomarker is predictive of the event time, the tail of the distribution has a higher/lower risk of event, and thus some individuals are removed from the population early on, resulting in a non-Gaussian distribution as time progresses). Additionally, the survival information was not used when fitting the longitudinal biomarker and fitting a new model at each event time may lead to different baseline biomarker values, which was an undesirable property (\citet{wulfsohn1997joint}). Thus joint models were introduced to simultaneously analyze longitudinal and event time processes assuming an association structure between these two outcomes, usually through correlated or shared random effects \citep{faucett1996simultaneously, wulfsohn1997joint, henderson2000joint}. Joint models address measurement error, missing data and informative right-censoring in the longitudinal process. They are useful to understand and quantify the association between the longitudinal marker and the risk of the terminal event, and to predict the risk of terminal event based on the history of the longitudinal marker. From an inferential point of view, they also reduce bias in parameter estimation and increase efficiency by utilizing all the available information from both the longitudinal and survival data simultaneously (e.g., to compare two treatment lines in a clinical trial), see \citet{rizopoulos2012joint}.\\

The joint modeling framework has been extended to account for multivariate longitudinal outcomes \citep{lin2002maximum, hickey2016joint} and competing risks of terminal events \citep{elashoff2008joint, hickey2018comparison} as usually encountered in health research. This implies however a substantial increase in the number of random effects to capture the correlation between all the processes in play and leads to a heavy computational burden, particularly due to the multidimensional integral over the random effects in the likelihood. Despite the frequent interest in multiple longitudinal endogenous markers measured over time in health-related studies, their simultaneous analysis in a comprehensive joint model has been limited so far by the available inference techniques and associated statistical software as specifically stated in \citet{hickey2016joint} and \citet{li2021flexible}. \citet{Mehdizadeh21} justifies the use of a two-stage inference approach for the joint analysis of a longitudinal marker with competing risks of events because ``the implementation of joint modeling involves complex calculations''. Also recently, \citet{MURRAY2022107438} discussed about the necessity to develop the multivariate joint modeling framework towards non-Gaussian longitudinal outcomes and competing risks of events.\\

The rapid evolution of computing resources and the development of statistical software made the estimation of a variety of joint models possible, initially using likelihood approaches \citep{rizopoulos2010jm, hickey2018joinerml}, with increasing reliance on Bayesian estimation strategies (e.g., Monte Carlo methods) \citep{JSSv072i07, rstan20}. Sampling-based methods are convenient for scaling up to more and more complex data. They have been proposed to fit multivariate joint models for longitudinal and survival data but always limited to relatively small samples and model specifications (i.e., small number of outcomes and simple regression models for each outcome), see for example \cite{ibrahim2004bayesian}. Moreover, these also have slow convergence properties. The Integrated Nested Laplace Approximation (INLA) algorithm has been introduced as an efficient and accurate alternative to Markov Chain Monte Carlo (MCMC) methods for Bayesian inference of joint models \citep{Rue09, JSSv100i02}. This algorithm requires to formulate the joint models as latent Gaussian models (LGM) and takes advantage of the model structure to provide fast approximations of the exact inference \citep{van2019joint}. The potential of INLA to provide fast and accurate estimation of joint models has been previously demonstrated in the specific context of a two-part joint model in comparison with maximum likelihood estimation with a Newton-like algorithm \citep{rustand2021bayesian}. This inference technique is thus a strong candidate for the deployment of joint models in increasingly complex settings encountered in health research where a large number of longitudinal markers and competing clinical events may be under study simultaneously.\\

The main contribution of this paper is the introduction of joint models for multivariate longitudinal markers with different distributions and competing risks of events, along with a fast and reliable inference technique applicable in practice. The finite sample properties of this inference technique are demonstrated in simulation studies and the flexibility of the approach is illustrated in a randomized placebo-controlled trial for the treatment of primary biliary cholangitis where the associations of 5 longitudinal markers of different nature with competing risks of death and liver transplantation are simultaneously assessed.\\

The paper is organized as follows. In Section \ref{metsec} we describe the multivariate joint model structure and the estimation strategy. In Section \ref{simsec} we present simulation studies to demonstrate the properties of this estimation strategy compared to the available alternatives. In Section \ref{appsec} we describe the application in primary biliary cholangitis before concluding with a discussion in Section \ref{dissec}.

\section{Methods}
\label{metsec}
\subsection{The joint model for multivariate longitudinal data and a multiple-cause terminal event}

Let $Y_{ijk}$ denote the $k^{th}$ $(k=1, ..., K)$ observed longitudinal measurement for subject $i$ $(i=1, ..., n)$ measured at time points $t_{ijk}$ with $j$ the occasion $(j=1, ..., n_{ik})$. The joint model for multivariate longitudinal data and a multiple-cause terminal event considers mixed effects models to fit the biomarkers' evolutions over time with link functions $g(\cdot)$ to handle various types of distributions, and a cause-specific proportional hazard model for the terminal event.\\

The model for the multivariate longitudinal part is defined as follows:
\[
\begin{array}{lc}
  \mathbb{E}[Y_{ijk}]=g^{-1}\left(\eta_{ik}(t_{ijk})\right)=g^{-1}\left(\bm X_{ik}(t_{ijk})^\top \bm \beta_k + \bm Z_{ik}(t_{ijk})^\top \bm b_{ik}\right) & \forall k=1,...,K
   \end{array}
\]
\noindent where $\eta_{ik}(t)$ is the linear predictor corresponding to individual $i$ and marker $k$ at time $t$, $\bm X_{ik}(t)$ and $\bm Z_{ik}(t)$ are vectors of covariates (possibly different for each marker $k$) associated with the fixed effects parameters $\bm \beta_k$ and the multivariate normal vector of random effects $\bm b_{i}$ which accounts for the intra-and-inter-marker correlation between the repeated measurements of an individual:
\begin{equation*}
\bm b_i = \begin{bmatrix}
\bm b_{i1}\\
\vdots\\
\bm b_{iK}
\end{bmatrix}
\sim MVN \left( \begin{bmatrix}
\bm{0} \\
\vdots \\
\bm{0} 
\end{bmatrix} , 
\begin{bmatrix}
\bm{\Sigma}_{\bm b_{i1}} & \cdots &\bm{\Sigma}_{\bm b_{i1} \bm b_{iK}} \\
\vdots & \ddots & \vdots\\
\bm{\Sigma}_{\bm b_{i1} \bm b_{iK}} & \cdots & \bm{\Sigma}_{\bm b_{iK}}
\end{bmatrix}\right).
\label{eq2}
\end{equation*}
The method and its implementation currently handle several distributions within the mixed model theory. This includes linear and lognormal models, generalized linear models (i.e., exponential family), proportional odds models for an ordinal outcome, zero-inflated models (Poisson, binomial, negative binomial and betabinomial) and two-part models for a semicontinuous outcome.\\

Additionally, let $T_i$ denote the positive continuous response variable that represents the elapsed time between the beginning of the follow-up and an event of interest for subject $i$. We can define the couple ($T_i^*, \delta_i$) where the actual observed time is $T_i^*=\textrm{min}(T_i, C_i)$ with $C_i$ the censoring time of individual $i$, and the indicator variable $\delta_i$ specifies the nature of the observed time, either cause $m$ ($m=1,...,M$) or censoring with $\delta_i = m \times I_{T_i<C_i}$. The model for the cause-specific survival part is defined as follows:
\[
\begin{array}{lc}
  \lambda_{im}(t)=\lambda_{0m}(t)\ \textrm{exp}\left(\bm W_{im}^\top(t) \bm \gamma_m +  h_m(\eta_{i1}(t), ..., \eta_{iK}(t)) \bm \varphi_m \right) & \forall m=1,...,M, \forall t \in \mathbb{R^+}.
   \end{array}
\]
The vector of possibly time-dependent covariates $\bm W_{im}(t)$ is associated with the risk of each cause $m$ of terminal event through regression parameters $\bm \gamma_m$ (see \citet{niekerk2021competing} for more details on competing risks joint models with \textbf{R-INLA}). The multivariate function $h_m(\eta_{i1}(t), ..., \eta_{iK}(t))$ defines the nature of the association with the risk of each cause $m$ of terminal event through the vector of parameters $\bm \varphi_m$. It can be specified as any linear combination of the linear predictors' components from the longitudinal submodels (therefore including shared random effects, current value and current slope parameterizations).\\

The baseline risk of event, $\lambda_{0m}(t)$, for each cause $m$, can be specified with parametric functions (e.g., exponential or Weibull) but it is usually preferred to avoid parametric assumptions on the shape of the baseline risk when it is unknown. We can define random-walks of order one or two corresponding to smooth spline functions based on first and second-order differences, respectively. The number of bins are not influential (as opposed to knots of other splines) since an increase in the number of bins only results in an estimate closer to the stochastic model. The second-order random-walk model provides a smoother spline compared to first-order since the smoothing is then done on the second-order. See \citet{martino2011approximate} and \citet{vanSankhya} for more details on the use of these random-walk models as Bayesian smoothing splines. We use these second-order random-walk smooth splines for the baseline hazard in the simulations and application sections (Sections \ref{simsec} and \ref{appsec}).

\subsection{Bayesian estimation with R-INLA}
\label{BayInf}
Let $\bm \lambda$ denote the parameters associated to the baseline risk functions $\lambda_{0m}(t)$, then the full vector of unknown parameters in the model is $\bm \theta_i=(\bm \beta_{1}, \cdots \bm \beta_{K}, \bm b_{i1} \cdots, \bm b_{iK}, \bm \lambda, \bm{\gamma}, \bm{\varphi})$. Let $\boldsymbol{D_i} \equiv \{T_i^*, \delta_i, Y_{ijk}: i=1, \cdots, n; j=1,\cdots, n_i, k=1, \cdots, K \}$ denote the observed variables. The Bayesian inference provides an estimation of the posterior distribution $p(\boldsymbol{\theta_i}|\boldsymbol{D_i})$ as defined by Bayes's theorem:
$$p(\bm \theta_i | \boldsymbol{D_i}) = \frac{p(\boldsymbol{D_i}| \bm \theta_i) p(\bm \theta_i)}{p(\boldsymbol{D_i})} \propto p(\boldsymbol{D_i} | \bm \theta_i) p(\bm \theta_i),$$
where $p(\boldsymbol{D_i} | \bm \theta_i)$ is the likelihood and $p(\bm \theta_i)$ is the joint prior. The marginal likelihood $p(\boldsymbol{D_i})=\int_{\bm \theta_i} p(\boldsymbol{D_i} | \bm \theta_i) p(\bm \theta_i) \mathrm{d}\bm \theta_i$ acts as a normalizing constant. The posterior distribution is often not analytically tractable and sampling-based methods like MCMC are used. Approximate methods like INLA provide exact approximations to the posterior at lower cost than sampling-based methods \citep{rue2017bayesian}.

\subsubsection{Likelihood function}

Assuming $\bm \theta_i =(\bm\phi, \bm b_i)$ where $\bm\phi=(\bm \beta_{1}, \cdots \bm \beta_{K}, \bm \lambda, \bm{\gamma}, \bm{\varphi})$ the vector of the fixed effects of the model and $\bm b_i = (\bm b_{i1} \cdots, \bm b_{iK})$ the random effects, we can derive the full likelihood of this model:
\begin{align*}
        p(\bm D_i|\bm\phi) = \int_{\bm b_{i}}  &\left[p(T_i^*, \delta_i | \bm \theta_i) \prod_{k=1}^{K} \prod_{j=1}^{n_i} p(Y_{ijk} | \bm \theta_i) p(\bm{b}_{i}) \right] \mathrm{d}\bm b_{i},
\end{align*}
where $p(T_i^*, \delta_i | \bm \theta_i) = \prod_{m=1}^M \lambda_{im}(T_i^*|\bm \theta_i)^{\mathbb{1}_{\delta_i=m}} \prod_{m=1}^M \exp\left( - \int_{0}^{T_i^*} \lambda_{im}(t|\bm \theta_i) \mathrm{d}t \right)$, $p(Y_{ijk} | \bm\theta_i)$ is the density of the $k^{th}$ longitudinal submodel and $p(\bm{b}_{i})$ is the density of the random effects.

\subsubsection{Integrated Nested Laplace Approximation}
In order to describe INLA's methodology, we need to formulate the joint model with a hierarchical structure with three layers where the first layer is the likelihood function of the observed data assumed conditionally independent. The second layer is formed by the latent Gaussian field $\bm u$ defined by a multivariate Gaussian distribution conditioned on the hyperparameters and the third layer corresponds to the prior distribution assigned to the hyperparameters $\bm \omega$, such that the vector of unknown parameters $\bm \theta$ from Section \ref{BayInf} is decomposed into two separate vectors, $\bm u$ and $\bm \omega$. For example the vector of fixed effects for longitudinal marker $k$, $\bm \beta_{k}$ have a prior defined as $\bm \beta_k \sim \mathcal{N}(0, \tau_{\beta_k} \bm I)$. The Gaussian parameters $\bm \beta_k$ belong to vector $\bm u$ and the corresponding hyperparameters $\tau_{\beta_k}$ belong to vector $\bm \omega$.

This model formulation is referred to as a Latent Gaussian Model. This general model formulation has computational advantages, particularly due to the sparsity of the precision matrix of the latent Gaussian field in the second layer of the LGM. Joint models can be formulated as latent Gaussian models and can therefore be fitted with INLA (\cite{martino2011approximate, van2019joint}). We assume an inverse-Wishart prior distribution for the covariance matrix of the random effects and Gaussian priors for the fixed effects. Finally, a penalizing complexity prior \citep{simpson2017penalising} is assumed for the precision parameter of the stochastic spline model for the log baseline hazard function.\\

The first step of INLA's methodology is an approximation of the marginal posterior distribution of the hyperparameters using the Laplace approximation:
$$p(\bm\omega|\pmb{D})\approx \frac{p(\bm\omega)p(\pmb{u}|\bm\omega)p(\pmb{D}|\pmb{u},\bm\omega)}{\tilde{p_G}(\pmb{u}|\bm\omega,\pmb{D})}|_{\bm u=\bm u^*(\bm\omega)},$$
where $\bm u^*(\bm\omega)$ denotes the mode of the full conditional for a given value of the vector of hyperparameters $\bm \omega$ and ${\tilde{p_G}(\pmb{u}|\bm\omega,\pmb{D})}$ is the Gaussian approximation obtained by matching the mode and curvature at the mode of the full joint density ${p(\pmb{u}|\bm\omega,\pmb{D})}$. The second step is an approximation of the conditional posterior distributions of the latent field:
		$$
		p(u_i|\pmb{\omega},\pmb{D}) \propto \frac{p(\pmb{u},\bm\omega|\pmb{D})}{p(\pmb{u}_{-i}|u_i,\bm\omega,\pmb{D})},
		$$
In the last step, a numerical integration is used to approximate the marginal posterior distributions of the latent field from the first two steps:
		$$
		p(u_i|\pmb{D})\approx \sum_{h=1}^{H}\tilde{p}(u_i|\bm\omega_h^*,\pmb{D})\tilde{p}(\bm\omega^*_h|\pmb{D})\Delta_h,
		$$
where the integration points $\{\bm\omega^*_1, ..., \bm\omega^*_H\}$ are selected from a rotation using polar coordinates and based on the density at these points and $\Delta_h$ are the corresponding weights. The approximation of the posterior marginal for each element of the latent field and each hyperparameter, using numerical integration, forms the ``integrated'' part of INLA algorithm while the first two steps above correspond to the ``nested Laplace'' approximation steps of INLA. In summary, INLA is a combination of nested numerical approximations described in \citet{NiekerkAvenue}. Originally, INLA used a series of two Laplace approximations, nested. The recent development suggests using a low-rank variational Bayes correction to match the full Laplace approximation with no significant additional computational cost to the first Gaussian approximation. The ``smart gradient" approach is used to more efficiently find the mode in the first step, see details in \citet{Esmail22}. Moreover, using numerical methods to deal with sparse precision matrix computations instead of dense covariance is crucial. Finally, \textbf{R-INLA} initially parameterized multivariate random effects with their precision and correlation but since the application of INLA to multivariate joint models involves random effects with high dimension, we developed a more stable way to deal with this in recent INLA versions by using the Cholesky parameterization which improves the computational stability and reduces the computational burden. More details about INLA are provided in Appendix A.\\

An alternative strategy referred to as empirical Bayes only uses the mode at step 1 (i.e., not the curvature at the mode that informs about uncertainty), which speeds up computations but cannot be considered as fully Bayesian. Pictured as a trade-off between frequentist and Bayesian estimation strategies, we show in our simulation studies (Section 3) that the empirical Bayes estimation strategy has frequentist properties as good as the full Bayesian strategy.\\

In the following we use \textbf{R-INLA} version INLA\_22.10.07 with the \textbf{PARDISO} library that provides a high performance computing environment with parallel computing support using OpenMP (\citet{JSSv100i02}). Additionally, we use the R package \textbf{INLAjoint} v23.07.12 (\citet{Rustand_INLAjoint}) that provides a user-friendly interface to fit joint models with \textbf{R-INLA}. See github.com/DenisRustand/MultivJoint for examples of codes that fit multivariate joint models with \textbf{INLAjoint}.

\subsection{Characteristics and limitations of the INLA technique}
\label{CharlimINLA}
In Bayesian inference, the gold standard would be an MCMC sampling (or some other Monte Carlo based method) running for an infinite number of iterations. For practical reasons, two types of approximation methods have been used: stochastic approximation methods, including primarily a finite number of MCMC runs, and deterministic approximation methods, including INLA amongst others (e.g., variational Bayes, expectation-propagation) which rely on analytical approximations. In the case of MCMC, the error is $\bm{\mathcal{O}_p}(T^{-1/2})$, where $T$ is the number of MCMC iterations (see \citet{Flegal08} for more details and assumptions). MCMC thus allows for a rough estimate at a minimal computational expense, and it is possible to achieve an arbitrary small error given sufficient computational time. In contrast, the error in INLA is relative and is determined by the combination of the prior and the likelihood (see Appendix B for an illustrative example). Because of the deterministic nature of the method, there are no tuning parameters to make the approximation error tend to zero at the cost of more computations. Its assessment would require running an MCMC for an infinite number of iterations, which is mostly not feasible, as with other approximation techniques. However, it is possible to adjust the precision of the approximations (for instance with tolerance parameters for the numerical integration necessary for the inference of the hyperparameters), which usually has little effect in non-extreme situations even though it may improve the stability of the results for complex models based on low-informative data.

INLA has been shown to be accurate and efficient for models with likelihood identifiability and/or suitable priors (\citet{Rue09, NiekerkAvenue, gaedke2022parallelized}). For ill-defined models, e.g., models with identifiability issues or models overly complicated compared to the available information, INLA can fail to accurately fit the model.  With MCMC, these models can be diagnosed by slow convergence or convergence issues. With INLA, some warnings can also be considered to indicate these pathogenic models. In Bayesian statistics, a posterior distribution very sensitive to prior specification will indicate a lack of data information available. We detail in Appendix C a procedure to evaluate the impact of priors on posterior distributions with INLA to possibly identify ill-defined models.

It should be noted that the INLA methodology has limitations, as discussed in \citet{Rue09}. First, it is specifically designed, and expected to perform well, for models that can be expressed as latent Gaussian models. Although this class of models includes most models used in longitudinal and survival analysis, there are exceptions, like non-Gaussian random effects and non-linear mixed models based on ordinary differential equations, that do not have an analytical solution. For Bayesian models that do not align with the LGM framework, MCMC sampling is often the only available option. 
Second, implementing new models in INLA demands a higher level of expertise compared to MCMC due to the intricate numerical optimization and programming involved. Statisticians who are comfortable with the lengthy computation time of an MCMC algorithm may favor this approach for LGMs. INLA merely offers a faster and reliable alternative for this specific class of models. Furthermore, INLA relies on large sparse matrix computations to approximate analytically the posterior distributions, and can require a substantial amount of memory for large models. INLA was initially designed to handle models with a low number of hyperparameters. Significant improvements have been made in this regard (\citet{gaedke2022parallelized}) but INLA may be less suitable for models with a large number of hyperparameters, even though we demonstrated the successful use of INLA with 150 hyperparameters in this work (these hyperparameters include random effects variance and covariance, residual error variance, association parameters, etc.).

\section{Simulations}
\label{simsec}
\subsection{Settings}

We undertook an extensive simulation study with two purposes: (i) to validate the inference approach based on INLA algorithm, (ii) to compare its performances with those obtained by two alternative existing inference techniques:
\begin{itemize}
\item Maximum likelihood estimation with Monte Carlo expectation maximization algorithm as implemented in the R package \textbf{joineRML} (version 0.4.5). This package can fit joint models of time-to-event data and multivariate longitudinal data in a frequentist framework using a recently introduced fast approximation that combines the expectation maximization algorithm to quasi-Monte Carlo \citep{hickey2018joinerml, philipson2020faster}. However, this method is restricted to Gaussian markers and the association structure between the longitudinal and survival part is limited to the linear combination of the random effects, i.e., individual deviation from the population mean. This parameterization is equivalent to the ``current value of the linear predictor'' in two situations: in the absence of covariates in the longitudinal part of the model, and when the covariates are also included in the survival part. In our simulation setting, the generation model did not include the covariates from the longitudinal part in the survival submodel. So, in the estimation procedure, we adjusted the survival part on the two covariates (binary and continuous) with
joineRML in order to have a comparable association across software. We used quasi-Monte Carlo with Sobol sequence and same options as described in \citet{philipson2020faster} for the simulations.
\item Bayesian inference via Markov Chain Monte Carlo sampling as implemented in the R package \textbf{rstanarm} (version 2.21.1) which relies on the Hamiltonian Monte Carlo algorithm implemented in \textbf{Stan} \citep{rstan20}. It can fit joint models of time-to-event data and multivariate longitudinal data with various types of distributions (i.e., generalized linear mixed effects models) but is limited to a maximum of 3 longitudinal outcomes (we limited our simulation studies to a maximum of 3 longitudinal markers for this reason). Alternatively, the R package \textbf{JMbayes2} (version 0.3-0) recently released, can fit joint models of time-to-event data and multivariate longitudinal data using MCMC implemented in C++ \citep{JMbayes222} with various options for the distribution of the longitudinal outcomes. It handles competing risks and multi-state models for survival outcomes.
\end{itemize}

\begin{figure}[t]
\centering
\includegraphics[scale=0.5]{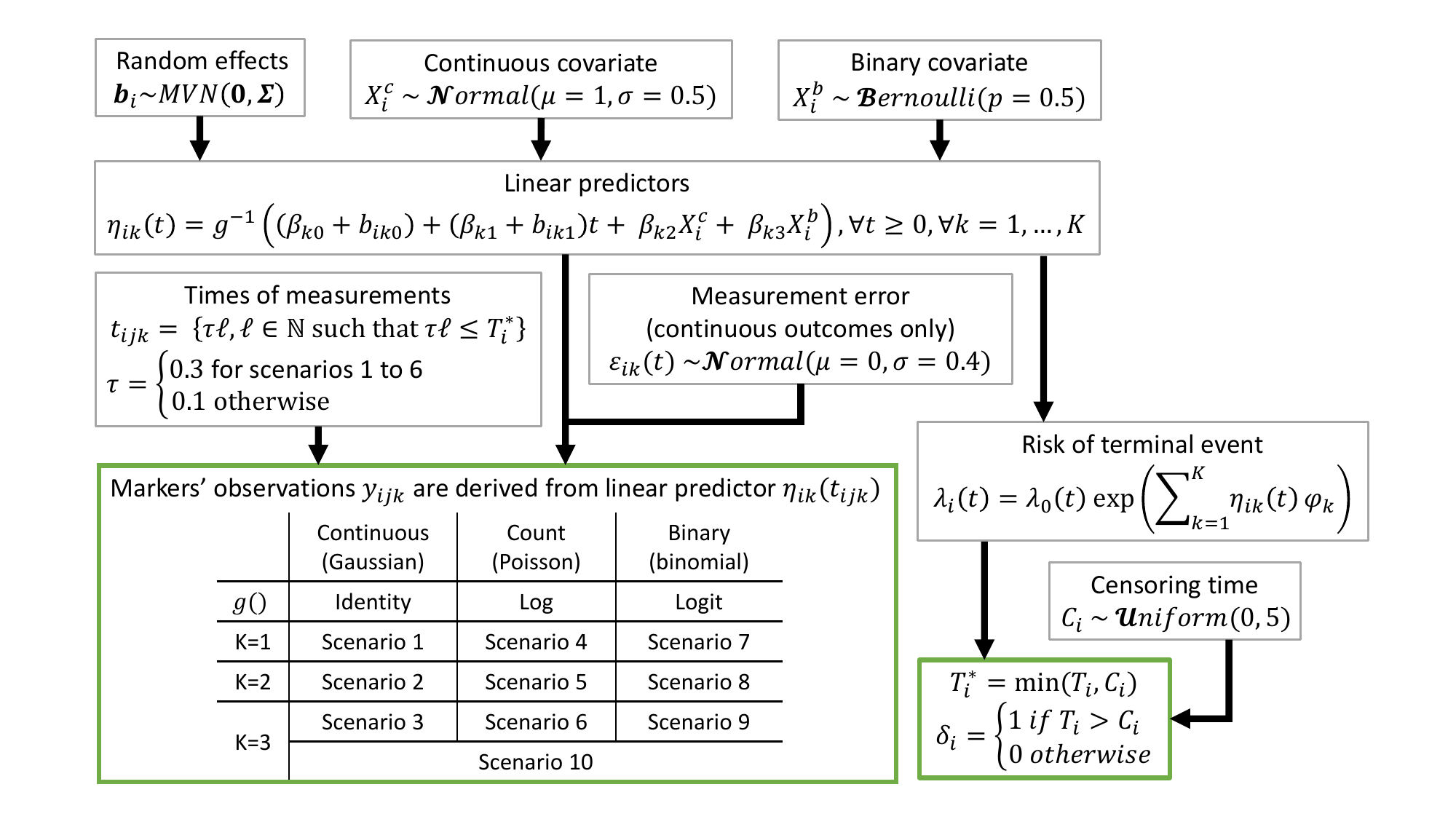}
\caption{Data generation settings for the simulation studies}
\label{Fig1}
\end{figure}

We are interested in the frequentist properties of the different estimation strategies and we therefore report the absolute bias, the standard deviation of the absolute bias and the frequentist coverage probability of the 95\% credible intervals for Bayesian inference and 95\% confidence intervals for the frequentist estimation implemented in \textbf{joineRML}. We use the same prior distributions for the fixed effects parameters of the Bayesian estimations, namely the default priors defined in \textbf{rstanarm} (Gaussian distributions with mean 0 and scale 2.5). In contrast, \textbf{JMbayes2} defines prior distributions using frequentist univariate regression models fitted separately for each outcome. The mean of the Gaussian prior is defined as the maximum likelihood estimate (MLE) and the precision is defined as the inverse of 10 times the variance of the estimate from the univariate model (only the prior for the association parameter matches the other Bayesian estimation strategies since it is not given by univariate models). Regarding the prior distributions for the multivariate Gaussian random effects variance and covariance parameters, \textbf{R-INLA} uses the inverse-Wishart prior distribution, \textbf{JMbayes2} uses gamma priors with mean defined as the MLE from univariate models while \textbf{rstanarm}'s default prior is the Lewandowski-Kurowicka-Joe (LKJ) distribution (\citet{LEWANDOWSKI20091989}), described as faster than alternative priors in the package documentation. Starting values for \textbf{R-INLA} are kept as default (i.e., non-informative) while \textbf{joineRML}, \textbf{JMbayes2} and \textbf{rstanarm} use MLE values from univariate models as initial values for both fixed and random effects (\textbf{joineRML} includes an additional step that fits a multivariate linear mixed effects sub-model). We note that these initial values provide an important advantage for \textbf{joineRML}, \textbf{JMbayes2} and \textbf{rstanarm} but we decided to keep them because this is part of the procedure when using these packages while \textbf{R-INLA} does not rely on submodels MLE values, although informative priors and initial values can also be considered. The baseline hazard is modeled with a different approach for each package: \textbf{R-INLA} defines Bayesian smoothing spline for the log baseline hazard with 15 bins, \textbf{rstanarm} uses cubic B-splines for the log baseline hazard with 2 internal knots, \textbf{JMbayes2} uses quadratic B-splines with 9 internal knots and the frequentist implementation from \textbf{joineRML} keeps the baseline hazard unspecified.\\

The design of simulation is fully described in Figure \ref{Fig1}. We considered 10 scenarios depending on the number of longitudinal markers (from 1 to 3) and the nature of the markers (Gaussian, Poisson or binomial). The distributions are chosen as the most common for each data type but \textbf{INLAjoint} also offers more complex and non-standard distribution choices. We considered linear trajectories for each marker, and the linear predictor of each marker at the time of risk assessment for the structure of association between the longitudinal and event processes as the common available association structure across software; for joineRML this requires to further adjust the survival model on the two covariates included in the marker trajectory. The time of event was generated by the methodology of Sylvestre and Abrahamowicz implemented in the R package PermAlgo \citep{sylvestre2008comparison}. The parameter values of the generation models were based on the real data analysis (Section \ref{appsec}). For each simulation scenario, we generated 300 datasets that included 500 individuals each where approximately 40\% of the sample observed the terminal event before the end of the follow-up, and each longitudinal marker had approximately 3000 total repeated measurements (min $\simeq 1$, max $\simeq 17$, median $\simeq 5$ per individual), except for simulation scenarios including at least one binary outcome. Models for continuous and count outcomes systematically included both a random intercept and a random slope while models for binary outcomes only included a random intercept. In addition, in scenarios including a binary outcome (Scenarios 7, 8, 9, 10), the number of repeated measures per individual was increased leading to approximately 8500 repeated measurements with min $\simeq 1$, max $\simeq 50$ and median $\simeq 13$ per individuals. In preliminary simulations, results were not satisfying (i.e., substantial bias and poor coverage for the variance-covariance of the random effects) regardless of the estimation strategy when using random intercept and slope as well as a limited number of repeated measurements per individual. Note that for scenario 10 (including a binary, a continuous and a count outcome), only the model for the binary outcome includes only a random intercept, i.e., the models for continuous and count outcomes include both random intercept and random slopes for a total of 5 correlated random effects.\\
 
We considered two estimation strategies for \textbf{R-INLA}, namely the empirical Bayes strategy and full Bayesian strategy (see Section \ref{metsec}). We also tested two different estimation strategies for \textbf{rstanarm}; the first one included 1 chain and 1000 MCMC iterations (including a warmup period of 500 iterations that was discarded) while the other one had default specifications (4 chains and 2000 iterations, including a warmup period of 1000 iterations that was discarded). The objective was to illustrate the impact of the choice of the number of chains and iterations on the parameter estimates and on the computation time. We kept the default specifications with \textbf{JMbayes2} since it provides a good trade-off between speed and accuracy (3 chains and 3500 iterations including a warmup period of 500 iterations that is discarded), the convergence of the chains was monitored in pre-studies (not shown); it was stable and consistent across scenarios. Note that the default was increased to 6500 iterations for binary outcomes but we finally kept 3500 iterations for those scenarios as the additional iterations were increasing the computation time for a negligible improvement in the results. All the computation times in the results are given with parallel computations over 15 CPUs (Intel Xeon Gold 6248 2.50GHz).\\

A replication script is available at github.com/DenisRustand/MultivJoint; it includes the generation and estimation codes for scenario 10 with 3 longitudinal markers of different natures.

\subsection{Results}

Since \textbf{joineRML} is restricted to Gaussian markers, we first report the results of the scenarios with Gaussian continuous outcomes only, and then report those with counts and binary outcomes.

\begin{table}
\centering
\caption{Simulations with $K$=3 continuous longitudinal markers}
\footnotesize
{\tabcolsep=1pt
\begin{tabular}{@{}lccccccccccccccccccccccccccccc@{}}
\hline
Approach: &  \multicolumn{3}{c}{R-INLA 1} & \multicolumn{3}{c}{R-INLA 2} & \multicolumn{3}{c}{joineRML*} & \multicolumn{3}{c}{JMbayes2} & \multicolumn{3}{c}{rstanarm 1} & \multicolumn{3}{c}{rstanarm 2}\\
 & \multicolumn{3}{c}{(Empirical Bayes)} & \multicolumn{3}{c}{(Full Bayesian)} & & & & & & & \multicolumn{3}{c}{(1 chain / 1000 iter.)} & \multicolumn{3}{c}{(4 chains / 2000 iter.)}\\
True value & Bias & (SD) & CP (\%) & Bias & (SD) & CP & Bias & (SD) & CP & Bias & (SD) & CP & Bias & (SD) & CP & Bias & (SD) & CP  \\
\hline
$\beta_{10}$=0.2 & 0 & (0.055) & 94 & 0 & (0.055) & 94 & -0.002 & (0.055) & 95 & 0 & (0.055) & 94 & -0.002 & (0.054) & 93 & 0.001 & (0.055) & 93 \\
 $\beta_{11}$=-0.1 & 0 & (0.023) & 96 & 0 & (0.023) & 96 & -0.007 & (0.023) & 97 & 0.001 & (0.023) & 96 & 0 & (0.024) & 96 & 0 & (0.024) & 96 \\
 $\beta_{12}$=0.1 & -0.004 & (0.046) & 92 & -0.004 & (0.046) & 92 & -0.004 & (0.046) & 93 & -0.004 & (0.046) & 92 & -0.002 & (0.046) & 91 & -0.003 & (0.046) & 92 \\
 $\beta_{13}$=-0.2 & 0.004 & (0.042) & 94 & 0.004 & (0.042) & 95 & 0.004 & (0.042) & 95 & 0.005 & (0.042) & 95 & 0.008 & (0.042) & 95 & 0.004 & (0.043) & 94 \\
 $\sigma_{\varepsilon 1}$=0.4 & -0.001 & (0.006) & 95 & -0.001 & (0.006) & 95 & 0 & (0.006) & 90 & 0 & (0.006) & 95 & 0.001 & (0.005) & 95 & 0.001 & (0.006) & 94 \\
 $\beta_{20}$=0.2 & -0.002 & (0.062) & 95 & -0.002 & (0.062) & 95 & -0.002 & (0.062) & 96 & -0.002 & (0.062) & 95 & 0.002 & (0.059) & 96 & -0.003 & (0.061) & 96 \\
 $\beta_{21}$=-0.1 & 0.001 & (0.031) & 94 & 0.001 & (0.031) & 94 & 0.016 & (0.031) & 91 & 0.001 & (0.031) & 94 & 0.002 & (0.03) & 93 & 0 & (0.031) & 94 \\
 $\beta_{22}$=0.1 & 0.002 & (0.052) & 95 & 0.002 & (0.052) & 94 & 0.002 & (0.052) & 94 & 0.003 & (0.052) & 94 & -0.001 & (0.05) & 96 & 0.004 & (0.051) & 95 \\
 $\beta_{23}$=-0.2 & -0.001 & (0.051) & 94 & -0.001 & (0.051) & 94 & -0.001 & (0.051) & 94 & -0.001 & (0.051) & 94 & -0.002 & (0.053) & 93 & -0.001 & (0.052) & 93 \\
 $\sigma_{\varepsilon 2}$=0.4 & 0 & (0.006) & 94 & 0 & (0.006) & 95 & -0.001 & (0.006) & 91 & 0 & (0.006) & 95 & 0 & (0.006) & 95 & 0 & (0.006) & 94 \\
 $\beta_{30}$=0.2 & -0.003 & (0.06) & 96 & -0.003 & (0.059) & 96 & -0.003 & (0.059) & 97 & -0.002 & (0.059) & 96 & -0.002 & (0.061) & 94 & -0.003 & (0.06) & 96 \\
 $\beta_{31}$=-0.1 & 0.003 & (0.024) & 97 & 0.003 & (0.024) & 97 & -0.007 & (0.023) & 96 & 0.002 & (0.024) & 98 & 0.003 & (0.024) & 97 & 0.003 & (0.023) & 98 \\
 $\beta_{32}$=0.1 & 0.006 & (0.047) & 95 & 0.006 & (0.047) & 95 & 0.006 & (0.048) & 96 & 0.006 & (0.047) & 94 & 0.007 & (0.049) & 94 & 0.007 & (0.048) & 95 \\
 $\beta_{33}$=-0.2 & 0.001 & (0.052) & 95 & 0.001 & (0.052) & 94 & 0 & (0.052) & 95 & 0 & (0.052) & 95 & -0.002 & (0.052) & 95 & 0 & (0.051) & 95 \\
 $\sigma_{\varepsilon 3}$=0.4 & -0.001 & (0.006) & 95 & -0.001 & (0.006) & 95 & 0 & (0.006) & 89 & 0 & (0.006) & 95 & 0 & (0.006) & 94 & 0.001 & (0.006) & 95 \\
 $\sigma_{b10}^2$=0.16 & 0.005 & (0.017) & 96 & 0.004 & (0.015) & 96 & -0.003 & (0.015) & 94 & 0.002 & (0.015) & 99 & 0.001 & (0.017) & 91 & 0.001 & (0.016) & 92 \\
 $\sigma_{b11}^2$=0.16 & 0.005 & (0.016) & 95 & 0.005 & (0.016) & 95 & -0.003 & (0.016) & 94 & -0.001 & (0.016) & 94 & 0.002 & (0.016) & 94 & 0.001 & (0.017) & 93 \\
 $\sigma_{b20}^2$=0.25 & 0.005 & (0.02) & 97 & 0.005 & (0.019) & 97 & 0 & (0.019) & 98 & 0.005 & (0.02) & 96 & 0.005 & (0.02) & 96 & 0.005 & (0.02) & 96 \\
 $\sigma_{b21}^2$=0.25 & 0.002 & (0.023) & 96 & 0.003 & (0.028) & 97 & -0.004 & (0.023) & 96 & 0.004 & (0.023) & 93 & 0.007 & (0.025) & 92 & 0.006 & (0.024) & 93 \\
 $\sigma_{b30}^2$=0.25 & 0.003 & (0.021) & 96 & 0.003 & (0.021) & 96 & -0.003 & (0.022) & 95 & 0.002 & (0.022) & 95 & 0.005 & (0.023) & 93 & 0.004 & (0.022) & 94 \\
 $\sigma_{b31}^2$=0.16 & 0.005 & (0.017) & 95 & 0.004 & (0.016) & 94 & -0.004 & (0.016) & 94 & 0.001 & (0.016) & 94 & 0.003 & (0.016) & 94 & 0.003 & (0.016) & 95 \\
 $\textrm{cov}_{b10,b11}$=0.08 & -0.006 & (0.012) & 92 & -0.006 & (0.012) & 91 & -0.001 & (0.011) & 96 & -0.003 & (0.011) & 95 & -0.002 & (0.011) & 94 & -0.001 & (0.011) & 94 \\
 $\textrm{cov}_{b10,b20}$=0.02 & 0.001 & (0.013) & 97 & 0.001 & (0.012) & 98 & 0 & (0.012) & 95 & 0.001 & (0.012) & 95 & 0 & (0.012) & 94 & 0 & (0.012) & 93 \\
 $\textrm{cov}_{b10,b21}$=0.04 & -0.003 & (0.014) & 96 & -0.003 & (0.014) & 96 & -0.002 & (0.014) & 96 & -0.002 & (0.013) & 95 & -0.002 & (0.014) & 93 & -0.002 & (0.014) & 94 \\
 $\textrm{cov}_{b10,b30}$=0 & -0.001 & (0.013) & 96 & -0.001 & (0.013) & 96 & 0 & (0.012) & 96 & 0 & (0.012) & 95 & 0 & (0.012) & 95 & 0 & (0.012) & 96 \\
 $\textrm{cov}_{b10,b31}$=-0.04 & 0.002 & (0.012) & 95 & 0.001 & (0.012) & 95 & 0.001 & (0.012) & 94 & 0.002 & (0.011) & 93 & 0.001 & (0.012) & 93 & 0.001 & (0.012) & 94 \\
 $\textrm{cov}_{b11,b20}$=0.04 & -0.003 & (0.014) & 94 & -0.004 & (0.014) & 94 & -0.002 & (0.014) & 94 & -0.002 & (0.013) & 94 & -0.001 & (0.014) & 92 & -0.001 & (0.014) & 93 \\
 $\textrm{cov}_{b11,b21}$=0 & 0.001 & (0.013) & 97 & 0.001 & (0.014) & 96 & -0.002 & (0.013) & 97 & -0.001 & (0.013) & 95 & -0.001 & (0.013) & 96 & -0.001 & (0.013) & 96 \\
 $\textrm{cov}_{b11,b30}$=-0.08 & 0.001 & (0.015) & 95 & 0.001 & (0.016) & 94 & -0.001 & (0.015) & 95 & 0.003 & (0.014) & 92 & 0 & (0.015) & 93 & 0 & (0.015) & 92 \\
 $\textrm{cov}_{b11,b31}$=-0.08 & 0.002 & (0.012) & 95 & 0.002 & (0.012) & 95 & 0 & (0.012) & 96 & 0.004 & (0.012) & 92 & 0.001 & (0.012) & 95 & 0.002 & (0.012) & 94 \\
 $\textrm{cov}_{b20,b21}$=0.1 & -0.002 & (0.016) & 97 & -0.001 & (0.017) & 97 & 0 & (0.016) & 98 & -0.002 & (0.015) & 96 & -0.001 & (0.014) & 98 & 0 & (0.015) & 97 \\
 $\textrm{cov}_{b20,b30}$=0.05 & 0.002 & (0.015) & 97 & 0.001 & (0.015) & 97 & 0 & (0.015) & 97 & 0.001 & (0.015) & 95 & 0.002 & (0.015) & 96 & 0.001 & (0.015) & 97 \\
 $\textrm{cov}_{b20,b31}$=0.02 & -0.001 & (0.013) & 97 & -0.002 & (0.013) & 97 & 0.001 & (0.013) & 98 & 0 & (0.012) & 98 & 0 & (0.013) & 97 & 0.001 & (0.013) & 98 \\
 $\textrm{cov}_{b21,b30}$=0.05 & 0 & (0.017) & 95 & 0 & (0.017) & 95 & 0.001 & (0.017) & 96 & -0.001 & (0.017) & 95 & 0.002 & (0.015) & 96 & 0.001 & (0.015) & 97 \\
 $\textrm{cov}_{b21,b31}$=-0.04 & 0.002 & (0.016) & 95 & 0.001 & (0.014) & 95 & 0.001 & (0.014) & 96 & 0.001 & (0.014) & 95 & -0.002 & (0.014) & 93 & 0 & (0.014) & 95 \\
 $\textrm{cov}_{b30,b31}$=0.1 & -0.003 & (0.014) & 96 & -0.004 & (0.014) & 95 & 0.003 & (0.014) & 95 & -0.003 & (0.014) & 95 & -0.001 & (0.015) & 94 & -0.001 & (0.014) & 95 \\
 $\varphi_1$=0.5 & 0.008 & (0.095) & 96 & 0.006 & (0.095) & 96 & 0.003 & (0.097) & 97 & 0.017 & (0.097) & 98 & 0.019 & (0.092) & 96 & 0.015 & (0.098) & 95 \\
 $\varphi_2$=-0.5 & -0.001 & (0.076) & 94 & -0.001 & (0.075) & 95 & -0.001 & (0.076) & 97 & -0.011 & (0.077) & 96 & -0.009 & (0.075) & 93 & -0.011 & (0.074) & 94 \\
 $\varphi_3$=0.5 & -0.003 & (0.087) & 95 & -0.004 & (0.087) & 95 & -0.008 & (0.088) & 98 & 0.006 & (0.089) & 96 & 0.011 & (0.09) & 95 & 0.01 & (0.088) & 96 \\
 Conv. rate & \multicolumn{3}{c}{1} & \multicolumn{3}{c}{1} & \multicolumn{3}{c}{1} & \multicolumn{3}{c}{1} & \multicolumn{3}{c}{0.65} & \multicolumn{3}{c}{0.83} \\
 Comp. time (sec.) & \multicolumn{3}{c}{53.38 (7.2)} & \multicolumn{3}{c}{88.75 (8.58)} & \multicolumn{3}{c}{51.98 (14.49)} & \multicolumn{3}{c}{166.41 (62.4)} & \multicolumn{3}{c}{1015.24 (405.83)} & \multicolumn{3}{c}{1247.33 (688.58)}\\
 \arrayrulecolor{black}\hline
 \multicolumn{19}{l}{*In joineRML, the two covariates were further included in the survival model to make the association structure comparable across software}
\end{tabular}}
\label{sim3Y}
\end{table}

\subsubsection{Gaussian longitudinal markers only}

The summary of the estimations provided by the two \textbf{R-INLA} strategies, the frequentist estimation with \textbf{joineRML} and the Bayesian estimations with \textbf{JMbayes2} and \textbf{rstanarm} are reported in Table 1 for 3 longitudinal Gaussian markers, and in Tables S1 and S2 of the supporting information for 1 and 2 markers, respectively. All the methods provided correct inference with negligible bias and correct coverage rate of the 95\% credible or confidence intervals. However, methods differed in terms of computation time and convergence rate. While \textbf{R-INLA}, \textbf{joineRML} and \textbf{JMbayes2} systematically converged, convergence issues appeared for \textbf{rstanarm} with the increasing number of markers (up to 65\% and 83\% in the scenario with 3 markers when running 1 and 4 chains, respectively).\\

\begin{figure}[!ht]
\centering
\includegraphics[scale=0.85]{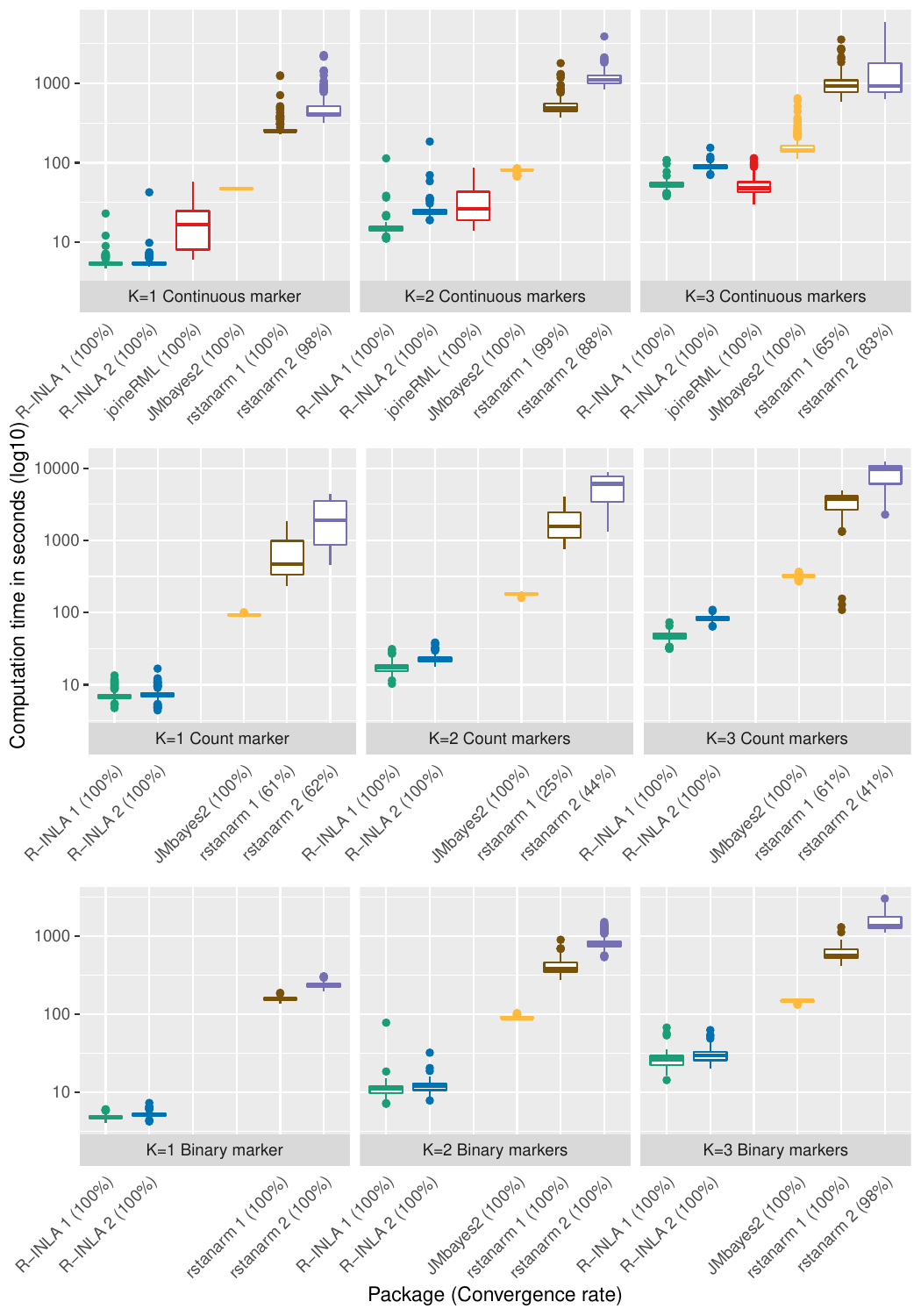}
\caption{Computation times and convergence rates for the 9 simulations scenarios with K=1,2 and 3 longitudinal markers with continuous, count and binary distributions.}
\label{Fig2}
\end{figure}

Computing times are summarized in Figure \ref{Fig2}. With 1 marker, \textbf{R-INLA} methodology presented the lowest computation time (5.52 and 5.65 seconds on average per dataset for the empirical Bayes strategy and full Bayesian strategy, respectively) followed by \textbf{joineRML} (19.09s. on average) and \textbf{JMbayes2} (47.09s. on average). When the number of markers increased, \textbf{joineRML} had the best scaling with similar computation times as the empirical Bayes version of \textbf{R-INLA} for 3 markers while the full Bayesian strategy of \textbf{R-INLA} was slightly longer. The computation time with \textbf{JMbayes2} for 2 and 3 markers was increased compared to \textbf{R-INLA} and \textbf{joineRML}. Finally, whatever the number of markers, \textbf{rstanarm} was much longer. With 3 markers, the 1 chain/1000 iterations strategy was approximately 19 times longer than the empirical Bayes INLA approach, and the 4 chain/2000 iterations strategy was approximately 23 times longer (ratios were even larger in smaller dimensions).

\begin{table}
\caption{Simulations with $K$=3 longitudinal markers with a mixture of distributions (continuous, count and binary)}
\footnotesize
\centering
{\tabcolsep=2.25pt
\begin{tabular}{@{}lccccccccccccccccccccccccccccc@{}}
\hline
Approach: &  \multicolumn{3}{c}{R-INLA 1} & \multicolumn{3}{c}{R-INLA 2} & \multicolumn{3}{c}{JMbayes2} & \multicolumn{3}{c}{rstanarm 1} & \multicolumn{3}{c}{rstanarm 2}\\
 & \multicolumn{3}{c}{(Empirical Bayes)} & \multicolumn{3}{c}{(Full Bayesian)} & & & & \multicolumn{3}{c}{(1 chain / 1000 iter.)} & \multicolumn{3}{c}{(4 chains / 2000 iter.)}\\
True value & Bias & (SD) & CP (\%) & Bias & (SD) & CP & Bias & (SD) & CP & Bias & (SD) & CP & Bias & (SD) & CP  \\
\hline
$\beta_{10}$=0.2 & 0.001 & (0.049) & 95 & 0.001 & (0.049) & 96 & 0.002 & (0.049) & 95 & 0.006 & (0.051) & 93 & -0.001 & (0.048) & 96 \\
 $\beta_{11}$=-0.1 & 0.003 & (0.019) & 96 & 0.003 & (0.019) & 95 & 0.002 & (0.019) & 95 & 0.003 & (0.021) & 93 & -0.001 & (0.019) & 94 \\
 $\beta_{12}$=0.1 & 0 & (0.04) & 95 & 0 & (0.04) & 94 & -0.001 & (0.04) & 94 & -0.004 & (0.041) & 93 & 0.004 & (0.039) & 97 \\
 $\beta_{13}$=-0.2 & -0.001 & (0.034) & 97 & 0 & (0.034) & 98 & 0 & (0.034) & 98 & -0.003 & (0.035) & 96 & -0.007 & (0.033) & 98 \\
 $\sigma_{\varepsilon 1}$=0.4 & 0 & (0.003) & 94 & 0 & (0.003) & 94 & 0 & (0.003) & 94 & 0.002 & (0.01) & 92 & 0.001 & (0.004) & 95 \\
 $\beta_{20}$=3 & -0.002 & (0.053) & 96 & -0.002 & (0.053) & 95 & 0.002 & (0.053) & 96 & 0.001 & (0.049) & 98 & -0.008 & (0.047) & 98 \\
 $\beta_{21}$=-0.1 & -0.002 & (0.02) & 96 & -0.002 & (0.02) & 96 & -0.001 & (0.02) & 96 & -0.002 & (0.019) & 97 & -0.003 & (0.019) & 98 \\
 $\beta_{22}$=0.1 & 0.002 & (0.043) & 96 & 0.002 & (0.043) & 95 & 0 & (0.043) & 96 & 0.001 & (0.041) & 95 & 0.005 & (0.041) & 95 \\
 $\beta_{23}$=-0.2 & 0.004 & (0.045) & 94 & 0.004 & (0.045) & 94 & 0.003 & (0.045) & 95 & 0.002 & (0.045) & 94 & 0.009 & (0.04) & 96 \\
 $\beta_{30}$=1 & -0.014 & (0.092) & 95 & -0.014 & (0.091) & 95 & -0.005 & (0.093) & 94 & 0.001 & (0.096) & 94 & 0.026 & (0.102) & 97 \\
 $\beta_{31}$=-1 & 0.008 & (0.031) & 93 & 0.009 & (0.031) & 93 & -0.001 & (0.031) & 95 & -0.001 & (0.032) & 94 & -0.007 & (0.031) & 96 \\
 $\beta_{32}$=1 & -0.006 & (0.074) & 96 & -0.007 & (0.073) & 95 & 0.004 & (0.075) & 97 & -0.002 & (0.076) & 94 & -0.008 & (0.071) & 98 \\
 $\beta_{33}$=-1 & 0.01 & (0.07) & 95 & 0.011 & (0.071) & 95 & -0.001 & (0.071) & 95 & -0.003 & (0.068) & 96 & -0.018 & (0.077) & 85 \\
 $\sigma_{b10}^2$=0.16 & 0.003 & (0.016) & 95 & 0.002 & (0.014) & 94 & -0.001 & (0.014) & 99 & -0.007 & (0.029) & 94 & -0.008 & (0.021) & 96 \\
 $\sigma_{b11}^2$=0.09 & 0.003 & (0.01) & 94 & 0.004 & (0.009) & 95 & 0 & (0.009) & 93 & 0.008 & (0.038) & 88 & 0.011 & (0.025) & 94 \\
 $\sigma_{b20}^2$=0.25 & 0.002 & (0.019) & 95 & 0.001 & (0.017) & 96 & 0.002 & (0.017) & 95 & 0.004 & (0.017) & 96 & 0.002 & (0.016) & 95 \\
 $\sigma_{b21}^2$=0.16 & 0.002 & (0.014) & 95 & 0.002 & (0.013) & 95 & 0.003 & (0.014) & 94 & 0.002 & (0.012) & 97 & 0.007 & (0.016) & 82 \\
 $\sigma_{b30}^2$=0.25 & -0.01 & (0.036) & 96 & -0.009 & (0.037) & 96 & 0 & (0.041) & 93 & 0.004 & (0.041) & 96 & 0.002 & (0.037) & 97 \\
 $\textrm{cov}_{b10,b11}$=0.03 & -0.002 & (0.009) & 96 & -0.004 & (0.009) & 94 & -0.001 & (0.007) & 97 & -0.001 & (0.01) & 96 & -0.003 & (0.009) & 96 \\
 $\textrm{cov}_{b10,b20}$=0.02 & 0 & (0.011) & 95 & 0 & (0.011) & 95 & 0 & (0.01) & 96 & -0.001 & (0.01) & 94 & -0.003 & (0.011) & 97 \\
 $\textrm{cov}_{b10,b21}$=0.04 & -0.001 & (0.01) & 96 & -0.001 & (0.01) & 96 & 0 & (0.009) & 96 & -0.002 & (0.011) & 94 & 0 & (0.009) & 97 \\
 $\textrm{cov}_{b10,b30}$=0 & 0 & (0.019) & 96 & -0.003 & (0.017) & 97 & 0.001 & (0.016) & 95 & -0.002 & (0.016) & 94 & -0.001 & (0.017) & 95 \\
 $\textrm{cov}_{b11,b20}$=0.03 & -0.001 & (0.01) & 96 & -0.001 & (0.01) & 96 & -0.001 & (0.009) & 96 & 0 & (0.011) & 94 & -0.002 & (0.01) & 97 \\
 $\textrm{cov}_{b11,b21}$=0 & 0 & (0.008) & 95 & 0 & (0.008) & 96 & 0 & (0.008) & 95 & 0.002 & (0.01) & 92 & 0 & (0.008) & 96 \\
 $\textrm{cov}_{b11,b30}$=-0.06 & 0.005 & (0.014) & 92 & 0.006 & (0.013) & 94 & 0.002 & (0.014) & 93 & 0 & (0.015) & 95 & -0.001 & (0.014) & 97 \\
 $\textrm{cov}_{b20,b21}$=0.08 & 0 & (0.012) & 95 & 0 & (0.012) & 95 & -0.001 & (0.011) & 94 & 0 & (0.011) & 96 & 0.002 & (0.011) & 96 \\
 $\textrm{cov}_{b20,b30}$=0.05 & -0.001 & (0.019) & 95 & -0.002 & (0.019) & 95 & -0.003 & (0.018) & 94 & -0.001 & (0.019) & 96 & 0.001 & (0.018) & 96 \\
 $\textrm{cov}_{b21,b30}$=0.04 & -0.001 & (0.017) & 95 & -0.001 & (0.017) & 96 & -0.003 & (0.016) & 94 & 0.009 & (0.019) & 92 & 0.011 & (0.018) & 95 \\
 $\varphi_1$=0.5 & -0.003 & (0.101) & 95 & -0.003 & (0.102) & 94 & 0 & (0.103) & 98 & 0.009 & (0.113) & 94 & 0.021 & (0.114) & 96 \\
 $\varphi_2$=-0.2 & 0.006 & (0.074) & 93 & 0.005 & (0.073) & 94 & 0.005 & (0.074) & 97 & 0.002 & (0.077) & 93 & 0.006 & (0.072) & 94 \\
 $\varphi_3$=0.3 & -0.001 & (0.094) & 94 & 0.001 & (0.092) & 94 & -0.004 & (0.092) & 94 & -0.006 & (0.093) & 94 & -0.003 & (0.087) & 95 \\
 Conv. rate & \multicolumn{3}{c}{1} & \multicolumn{3}{c}{1} & \multicolumn{3}{c}{1} & \multicolumn{3}{c}{0.51} & \multicolumn{3}{c}{0.64} & \\
 Comp. time (sec.) & \multicolumn{3}{c}{43.55 (5.58)} & \multicolumn{3}{c}{69.8 (44.59)} & \multicolumn{3}{c}{295.73 (20.22)} & \multicolumn{3}{c}{3766.41 (1429.06)} & \multicolumn{3}{c}{10861.22 (3735.02)}\\
 \arrayrulecolor{black}\hline
\end{tabular}}
\label{simGLM3Y}
\end{table}

\subsubsection{Non-Gaussian markers}

Results of scenarios with $k=1, 2$ and $3$ count longitudinal markers are presented in Tables S3, S4 and S5 of the supporting information, respectively. Results of scenarios with $k=1, 2$ and $3$ binary longitudinal markers are presented in Tables S6, S7 and S8 of the supporting information, respectively. Finally results of scenario 10 with one continuous marker, one count marker and one binary marker are summarized in Table \ref{simGLM3Y}.\\

From these 7 scenarios, the two strategies of INLA systematically provided correct inference with 100\% convergence, no bias, and coverage probabilities close to the nominal value. The MCMC strategy of \textbf{JMbayes2} also provided unbiased estimates with standard deviations that matched with \textbf{R-INLA} and coverage probabilities close to the nominal values for scenarios with counts longitudinal markers and the last scenario considering a mixture of markers. When considering only binary longitudinal markers, \textbf{JMbayes2} was not able to fit the scenario with only one marker because it requires to have at least two random effects in the model (while this scenario only included a random intercept). Additionally, with two and three binary longitudinal markers, \textbf{JMbayes2} lead to a higher bias compared to \textbf{R-INLA} and \textbf{rstanarm} for the association parameters. Note that we reduced \textbf{JMbayes2}'s default number of iterations for scenarios with binary outcomes to match the number of iterations of other scenarios. When using the default number of iterations (i.e., 6500) this bias was slightly reduced but remained higher compared to \textbf{R-INLA} and \textbf{rstanarm} (e.g., for $\varphi_1=0.3$, bias$=0.296$ with 3500 iterations (Table S8) and bias$=0.227$ with 6500 iterations for a computation time that increased from 147s. per model to 262s. per model (not shown)). The two MCMC strategies of \textbf{rstanarm} provided no bias and coverage probabilities close to the nominal value in the scenarios with only binary outcomes, and in the case of a mixture of markers, two cases where more repeated information had been considered. In contrast, with count data, the performances were poor with a small rate of convergence (between 41\% and 62\% for the 4 chains / 2000 iterations, between 25\% and 61\% for 1 chain / 1000 iterations), a higher bias and standard deviations together with low coverage probabilities, particularly with 1 chain and 1000 iterations.\\

Regarding computation time (Figure \ref{Fig2} for all counts and all binary outcomes, and Figure S1 for the mixed outcomes), as already observed with Gaussian markers, the two strategies of INLA were much faster than the MCMC strategies. For instance, when considering a mixture of markers, the computation time with \textbf{R-INLA} was less than one minute on average with empirical Bayes strategy (43.55s.) and slightly over one minute (69.8s.) with full Bayesian strategy while \textbf{JMbayes2} fitted one model in 5 minutes on average (295.73s.), \textbf{rstanarm} fitted a model on average in one hour (3766s.) with 1 chain and 1000 iteration and three hours (10861s.) with 4 chains and 2000 iterations.

The computation times were also between 40 and 300 times smaller with \textbf{R-INLA} compared to \textbf{rstanarm} and between 3 and 13 times smaller with \textbf{R-INLA} compared to \textbf{JMbayes2} in the count case, and between 20 and 75 times smaller with \textbf{R-INLA} compared to \textbf{rstanarm} and between 5 and 8 times smaller with \textbf{R-INLA} compared to \textbf{JMbayes2} in the binary case. Of note, the error messages following non-convergence with \textbf{rstanarm} indicated that running the chains for more iterations may help to meet the convergence criteria. However, the computation time was already so high that the most complex scenario (i.e., 3 markers with different distributions) required around 37 days straight to fit the 300 datasets with 4 chains and 2000 iterations despite parallel computing.

\section{Application to PBC}
\label{appsec}
\subsection{Description}
We illustrate the flexibility of our inference approach using \textbf{R-INLA} with an application to the study of primary biliary cholangitis (PBC), a rare chronic liver disease that often causes death. We leveraged the longitudinal data from 312 PBC patients followed at the Mayo Clinic between 1974 and 1988 who received either a placebo or D-penicillamine \citep{murtaugh1994primary}. These data are publicly available in several software including the R package \textbf{JM} \citep{rizopoulos2010jm}. During the follow-up, 140 patients died and 29 patients received liver transplantation which we consider here as a competing event of death. In addition, repeated measures of various longitudinal markers potentially associated with the disease progression were collected. Among them, we considered:
\begin{itemize}
\item 4 continuous markers: serum bilirubin (in mg/dl, log scale), serum aspartate aminotransferase (in U/ml, log scale), serum albumin (in gm/dl) and prothrombin time in seconds, 
\item 2 count markers: platelet (per cubic ml / 1000) and alkaline phosphatase (in U/liter)
\item 3 binary markers: spiders (presence of blood vessel malformations in the skin), ascites (presence of abnormal accumulation of fluid in the abdomen) and hepatomegaly (presence of enlarged liver).
\end{itemize}

\noindent The number of individual repeated measurements for these markers was 6.2 on average (minimum 1, maximum 16, median 5).

\subsection{Objective and strategy of analysis}

Our objective was to evaluate the association between the 9 longitudinal markers of progression and the competing risks of death and liver transplantation. We thus built a joint model for analyzing simultaneously the candidate markers and the 2 causes of event. The final joint model was built step-by-step. We first considered each marker separately and fitted joint models for one longitudinal marker and 2 competing causes of event in order to define the appropriate shape of the marker trajectory, choosing between linear and natural cubic splines with 2 knots at quantiles 0.33 (1 year of follow-up) and 0.66 (4 years of follow-up). The models systematically included a random intercept and random slopes on each function of time (i.e., two or four correlated random effects for linear or splines, respectively) as well as treatment (placebo or D-penicillamine) and its interactions with time functions. Then, we defined the structure of association with the two causes of event, choosing between current level (i.e. shared linear predictor) and current slope (or both). At this stage, we dropped the marker alkaline because it was not found associated with any cause of event and we also dropped prothrombin, ascites and hepatomegaly because they were giving unstable results (i.e., vague posterior distribution that mostly reflected the non-informative prior). We then built the final model including simultaneously the longitudinal markers found associated with the events in the previous step. Based on this strategy, the final model included 5 longitudinal markers, 3 continuous, 1 count and 1 binary. It was defined as follows for any time $t$:

\[
\left\{
\setlength\arraycolsep{0pt}
\begin{array}{ c @{{}{}} l  @{{}{}} r }
\log(Y_{i1}(t)) &= \eta_{i1}(t) + \varepsilon_{i1}(t) & \hspace{-6cm} \textbf{\textit{(bilirubin - lognormal)}}\\
&=(\beta_{10} + b_{i10})+\beta_{11}X_{i}+(\beta_{12}+b_{i11})\textrm{NS}_1(t)+(\beta_{13}+b_{i12})\textrm{NS}_2(t)+\\
& \ \ \ \hspace{0.05cm} (\beta_{14}+b_{i13})\textrm{NS}_3(t)+\beta_{15}X_{i}\textrm{NS}_1(t)+\beta_{16}X_{i}\textrm{NS}_2(t)+\beta_{17}X_{i}\textrm{NS}_3(t) + \varepsilon_{i1}(t) \\ \ \\

\log(Y_{i2}(t)) &= \eta_{i2}(t) + \varepsilon_{i2}(t) & \hspace{-6cm} \textbf{\textit{(aspartate aminotransferase - lognormal)}}\\
&= (\beta_{20} + b_{i20})+\beta_{21}X_{i}+(\beta_{22}+b_{i21})\textrm{NS}_1(t)+(\beta_{23}+b_{i22})\textrm{NS}_2(t)+\\
& \ \ \ \hspace{0.05cm} (\beta_{24}+b_{i23})\textrm{NS}_3(t)+\beta_{25}X_{i}\textrm{NS}_1(t)+\beta_{26}X_{i}\textrm{NS}_2(t)+\beta_{27}X_{i}\textrm{NS}_3(t) + \varepsilon_{i2}(t) \\ \ \\

Y_{i3}(t)&=\eta_{i3}(t) + \varepsilon_{i3}(t) & \hspace{-6cm} \textbf{\textit{(albumin - normal)}}\\
&= (\beta_{30} + b_{i30})+\beta_{31}X_{i}+(\beta_{32}+b_{i31})t+\beta_{33}X_{i}t+ \varepsilon_{i3}(t)\\ \ \\

\log(E[Y_{i4}(t)])&=\eta_{i4}(t) & \hspace{-6cm} \textbf{\textit{(platelet - Poisson)}}\\
&= (\beta_{40} + b_{i40})+\beta_{41}X_{i}+(\beta_{42}+b_{i41})\textrm{NS}_1(t)+(\beta_{43}+b_{i42})\textrm{NS}_2(t)+\\
& \ \ \ \hspace{0.05cm} (\beta_{44}+b_{i43})\textrm{NS}_3(t)+\beta_{45}X_{i}\textrm{NS}_1(t)+\beta_{46}X_{i}\textrm{NS}_2(t)+\beta_{47}X_{i}\textrm{NS}_3(t) \\ \ \\

\textrm{logit}(E[Y_{i5}(t)])&=\eta_{i5}(t) & \hspace{-6cm} \textbf{\textit{(spiders - binomial)}}\\
&= (\beta_{50} +b_{i50})+\beta_{51}X_{i}+(\beta_{52}+b_{i51})t+\beta_{53}X_{i}t \\ \ \\

\lambda_{i1}(t)&=\lambda_{01}(t)\ \textrm{exp}\left(\gamma_1 X_i + \eta_{i1}(t)\varphi_1 + \eta_{i1}'(t)\varphi_3 + \eta_{i2}(t)\varphi_4 \hspace{0.05cm} + \right. & \hspace{-6cm} \textbf{\textit{(death risk)}}\\
& \hspace{3.25cm} \left.\eta_{i3}(t)\varphi_5 + \eta_{i4}(t)\varphi_7 + \eta_{i5}(t)\varphi_9 \right) \\ \ \\

\lambda_{i2}(t)&=\lambda_{02}(t)\ \textrm{exp}\left(\gamma_2 X_i + \eta_{i1}(t)\varphi_2 + \eta_{i3}(t)\varphi_6 + \eta_{i4}(t)\varphi_8 \right) & \hspace{-6cm} \textbf{\textit{(transplantation risk)}}
\end{array}
\right.
\]

The independent Gaussian measurement errors for the first three markers are captured by $\varepsilon_{ik}(t)$, variable $X_i$ corresponds to the treatment received (placebo vs. D-penicillamine) and $\textrm{NS}_1(t), \textrm{NS}_2(t), \textrm{NS}_3(t)$ are the natural cubic splines with internal knots at 1 and 4 years. For the main analysis, we assumed correlated random effects within each marker but independent random effects across markers to avoid too many covariance parameters, resulting in 16 random effects and 20 covariance parameters. We estimated the model with the full variance-covariance matrix of the random effects (i.e., with 120 covariance parameters) in a sensitivity analysis. Compared to independent univariate joint models, the multivariate model has the asset of allowing the estimation of markers associations that are adjusted for the other markers, which is an essential feature, particularly in case of confounding as longitudinal markers can capture some individual heterogeneity.

\subsection{Results}
\begin{figure}[!ht]
\centering
\includegraphics[scale=0.8]{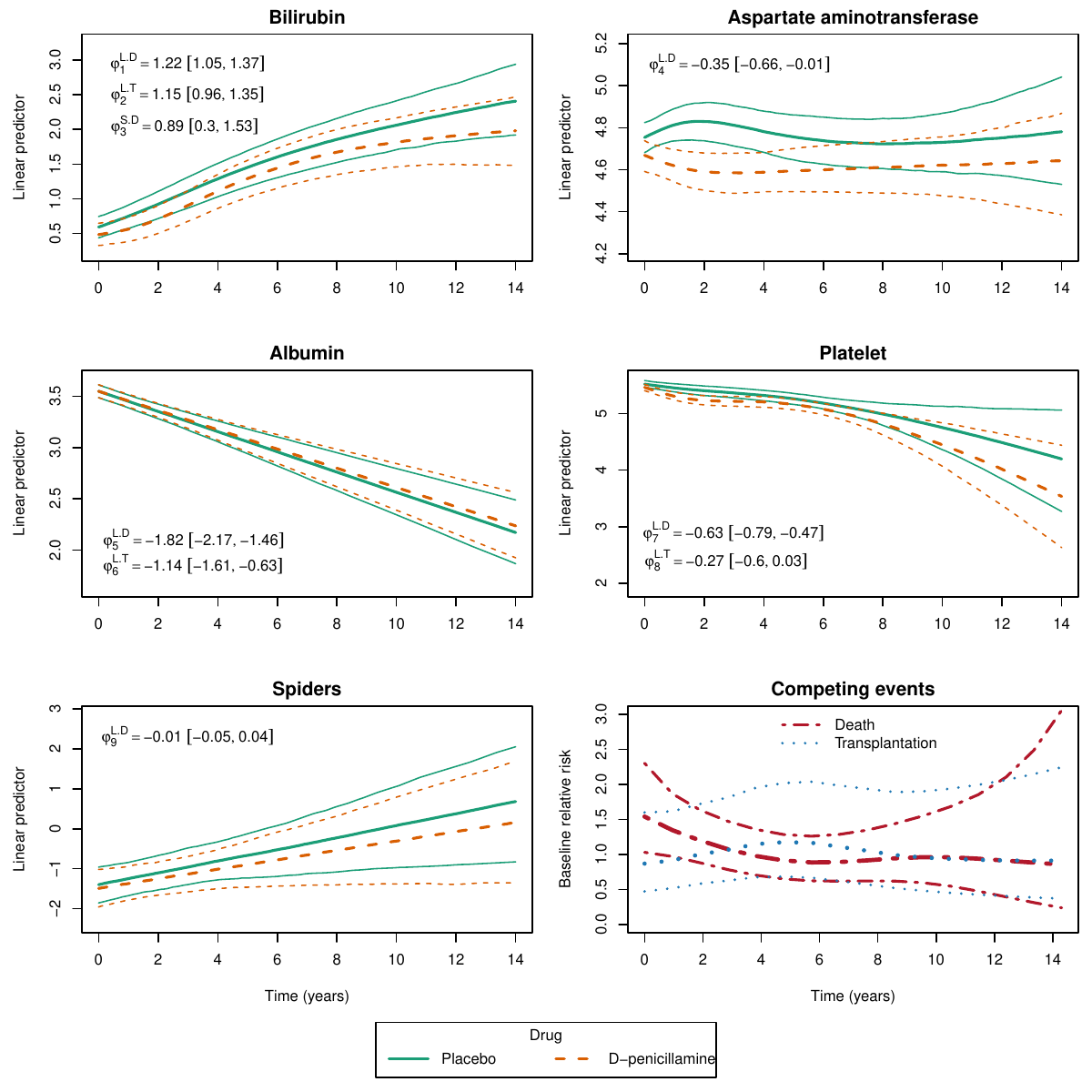}
\caption{Average linear predictor trajectories for each longitudinal marker and baseline risk curves. The 95\% bands for the uncertainty of the marker's trajectory is obtained by sampling from the joint posterior distributions (1000 samples). The association parameters between each longitudinal marker and the survival submodels are given in each plot by $\varphi^{*.*}$, where the first letter is ``L'' for current level of the linear predictor and ``S'' for current slope while the second letter corresponds to ``D'' for the effect on the risk of death and ``T'' for the risk of transplantation.}
\label{Fig3}
\end{figure}

The parameter estimates as well as their standard deviations and credible intervals are given in Table S9 of the supporting information. We illustrate the results with a plot of the average linear predictor trajectory conditional on treatment for each marker and the baseline risk of death and transplantation curves in Figure \ref{Fig3}. The computation time for this model (with the same resources as in the simulation studies) was 276 seconds. When relaxing the independence assumption between the inter-marker random effects (estimating the 120 covariance parameters), the computation time was substantially larger (2385 seconds) but the fixed effects and association parameters (reported in Table S10) were similar to those of the constrained model (Table S9). This illustrates that, in this example, modeling the correlation between markers does not seem to improve the model fit while it increases complexity by adding a lot of parameters.\\

Our results showed that the linear predictor describing the level of serum bilirubin increased but with a rate of change that diminished over time while platelet counts slowly decreased but with an increasing rate of change over time. The evolution of aspartate aminotransferase levels depended on the drug received; for patients receiving the placebo drug, it increased at the beginning of the follow-up and then stabilized while for patients receiving D-penicillamine, it slightly reduced in the early follow-up and then also stabilized. Finally, we observed an overall decrease in albumin concentration and an increase in the spiders's levels.\\

Regarding the association with clinical endpoints, our results exhibited a strong association between the log serum bilirubin and the risk of death through both its current level (log risk of death increased by 1.22 (95\% CI 1.05, 1.37) for each unit increase of the current value), and its current slope (0.89 (95\% CI 0.30, 1.53) increase of the log risk of death for each unit increase in the current slope of the log serum bilirubin) adjusted for treatment and all the other markers. Moreover, a unit increase in this biomarker's current value was associated to a 1.15 (95\% CI 0.96, 1.36) increase in the log risk of transplantation. The level of aspartate aminotransferase was slightly associated with decreased risks of death (association with log risk of death of -0.35, 95\% CI -0.66, -0.01). There was also a strong negative association between the current level of albumin and both the risk of death (-1.82 (95\% CI -2.17, -1.46) decrease in the log risk of death for a one unit increase) and the risk of transplantation ( -1.14 (95\% CI -1.61, -0.63) decrease in the log risk of transplantation for a one unit increase). After adjustment for treatment and the other markers, a higher current level of platelet counts was associated with a decreased risk of death (association with log risk of death of -0.63, 95\% CI -0.79, -0.47). Finally, the marker spiders was not found associated with the two causes of events, indicating that the association observed in the univariate model was fully explained by other markers included in the full model.

Our results regarding the treatment are consistent with the literature. For example, a meta-analysis evaluating the effect of placebo vs. D-penicillamine in randomized-clinical trials of primary biliary cholangitis patients by \citet{gong2006systematic} found no effect of this drug on the risk of death and the risk of transplantation. Also this meta-analysis found an association between D-penicillamine and a decrease of serum alanine aminotransferase, which is consistent with our finding where D-penicillamine is associated with a decrease of serum aspartate aminotransferase in early follow-up (serum alanine aminotransferase and aspartate aminotransferase are both enzymes that indicate liver damage). The R code for this application with \textbf{INLAjoint} package is available at github.com/DenisRustand/MultivJoint (``PBC\_INLA.R'').

\section{Discussion}
\label{dissec}

In this paper, we proposed a fast and accurate inference approach that makes it possible to fit flexible joint models for multivariate longitudinal and survival data. Based on the Integrated Nested Laplace Approximation Bayesian algorithm implemented in the R package \textbf{R-INLA}, it alleviates the computational burden of iterative estimation strategies implemented in classical software (i.e., maximum likelihood estimation or Bayesian inference with MCMC sampling), and thus allows the estimation of multivariate joint models with much fewer restrictions as illustrated in our simulation studies and application to primary biliary cholangitis. Several models were developed for the analysis of this dataset but they were often limited to one or few markers, did not account for the competing risks of events, or used a different approach than joint modeling to reduce the complexity (e.g., \citet{philipson2020faster, devaux2021individual, andrinopoulou2021reflections, Hughes21, MURRAY2022107438}).\\

We demonstrated that the estimation of this class of joint models with \textbf{R-INLA} is reliable and faster than any alternative estimation strategy (very good inference properties, computation time and convergence rate) currently implemented in software. Compared to alternative joint modeling software and inference techniques, the estimation of multivariate joint models with \textbf{R-INLA} has several other assets. There is no limitation for the number of longitudinal markers that can be included in the joint model and various distributions are available to describe the markers' distributions. For instance, in the longitudinal part, it handles zero-inflated models, two-part model or proportional odds model that are often of interest in health research (but currently not available in joint modeling software). It is also possible to consider competing risks of events as illustrated in the application (currently not possible with \textbf{joineRML} and \textbf{rstanarm}). Additionally, beyond competing risks \textbf{R-INLA} can handle various survival models in the context of joint modeling, including frailty models, multi-state models and mixture cure models. Finally, several parameterizations for the link between longitudinal and survival outcomes are available with \textbf{R-INLA}: shared random effects, current value and current slope of the biomarker or any linear combination of the biomarkers' linear predictors; and future extensions may allow for a non-linear effect of the biomarker on the risk of event. Since the \textbf{R-INLA} package is developed to fit a wide variety of statistical models beyond joint models, we used the user-friendly interface implemented in the R package \textbf{INLAjoint} that facilitates the use of INLA for joint modeling, as illustrated in the R codes ``MultivJoint.R'' and ``PBC\_INLA.R'' available at github.com/DenisRustand/MultivJoint that replicates the model fit for a simulated dataset and for the application Section of this paper, respectively.\\

There are however limitations with the INLA methodology. In addition to the general ones detailed in Section \ref{CharlimINLA}, some specifically concern the joint models. Because this inference technique is designed for models that can be expressed as latent Gaussian models, some nonlinear mixed submodels can not be considered. In addition, although the number of random effects is not limited, the number of correlated random effects is currently limited to 20 although this limit is practical and could be extended if necessary. As illustrated in the application, it does not preclude the inclusion of multiple independent groups of correlated random effects in a model. Furtermore, there is often interest in deriving individual predictions from the fitted model. This is not specifically implemented yet but the Bayesian properties of the model and the function available to sample from the joint posterior distribution make it easy to compute any prediction based on the model's output.\\

Our simulation studies mimic the design proposed in the function \textit{simData()} of the R package \textbf{joineRML} (see \citet{hickey2018joinerml}) but we extended it to non-Gaussian outcomes. Therefore the design is not specifically chosen to fit with \textbf{R-INLA} but rather based on previous simulations for multivariate joint models and limited by the range of models \textbf{joineRML} and \textbf{rstanarm} can fit. Our comparison of the available methodologies to fit multivariate joint models suggests that the iterative algorithms reach some limitations when fitting complex models, compared to the Bayesian approach implemented in \textbf{R-INLA}. In practice, INLA is specifically designed to fit latent Gaussian models and works well for any model that can be expressed as an LGM, it has the well-known strengths of the Bayesian framework for small data applications and its efficient computations make it scale well for complex models and data-rich applications. For the MCMC approach, our simulations show that \textbf{JMbayes2} has much better properties compared to \textbf{rstanarm}, both in terms of frequentist properties and computation time. It is the best alternative to \textbf{R-INLA} available at the moment since it is able to fit models with similar features for a moderate increase in computation time. Nevertheless, for complex models, a very high number of iterations may be required. For instance, in the application model with full covariance matrix, after 100 000 iterations in \textbf{JMbayes2} (CPU time of 10 hours when \textbf{R-INLA} procedure for the same model took 40 minutes), some of the parameters still had a Rhat$>$1.10 which suggests additional iterations are still required. 

The comparison between the software implementations in the simulation studies was not totally fair because the prior distributions across Bayesian strategies were not perfectly matching and because \textbf{joineRML}, \textbf{JMbayes2} and \textbf{rstanarm} used MLE from univariate submodels to define initial values and \textbf{JMbayes2} also used those MLE to define some of the prior distributions, note that it is possible to use the same non-informative priors with \textbf{JMbayes2} as with \textbf{rstanarm} and \textbf{R-INLA}. However, the purpose of our simulations was to illustrate the properties of the estimations strategies as they are meant to be used and therefore we decided to keep those differences to prevent software's misuse in the context of our comparison with \textbf{R-INLA}.\\

Until now, joint modeling software has only used iterative optimization (see for instance \citet{papageorgiou2019overview, furgal2019review} and \citet{alsefri2020bayesian}). It induces  limitations in the applicability of joint modeling. With a deterministic inference strategy, INLA offers a promising alternative for the development and application of joint models for multivariate longitudinal data and survival times as highlighted in this work.

\section*{Funding}
C\'ecile Proust-Lima's work was supported by the French National Research Agency (ANR) project JMECR (ANR-21-CE36-0013-01)

\bibliography{biblio}
\bibliographystyle{apalike} % <== and here you specify the style for your bibliography

\newgeometry{top=3cm, bottom=3cm, left=1.5cm, right=1.5cm}
%\hskip-2cm
\newpage
\section*{Supporting information}

\begin{figure}[ht]
\centering
\includegraphics[scale=0.7]{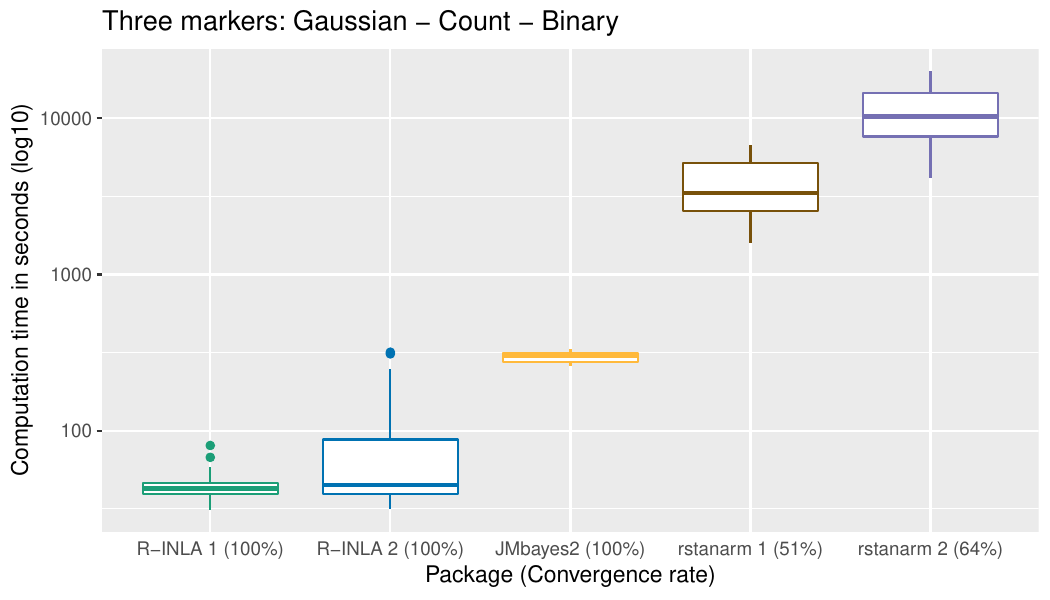}
\caption*{Figure S1: Computation times and convergence rates for simulations scenario 10 with a mixture of distributions (continuous, count and binary).}
\label{FigS1}
\end{figure}

\begin{figure}[ht]
\centering
\hspace{-1cm}
\includegraphics[scale=0.7]{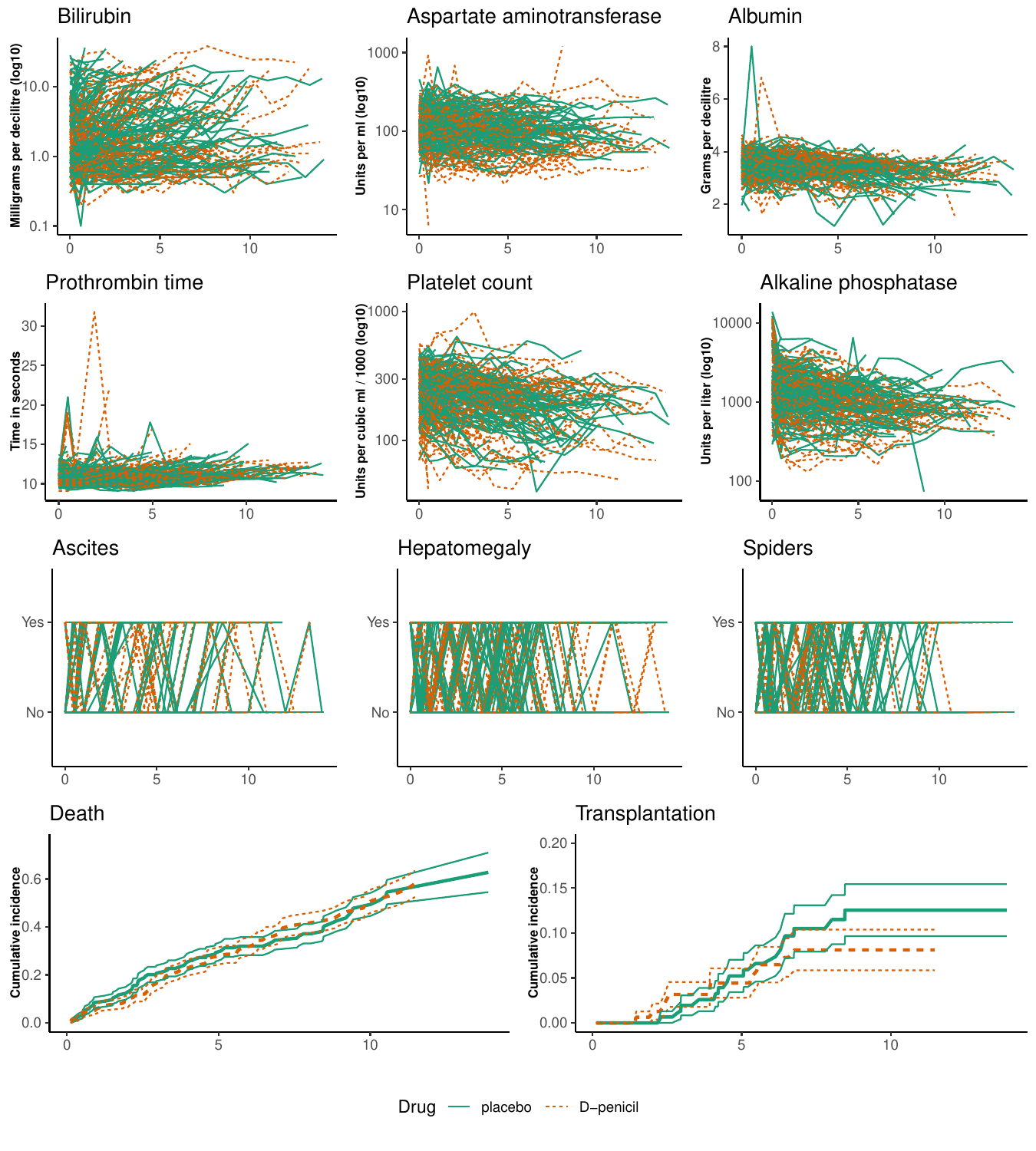}
\caption*{Figure S2: Description of the 9 longitudinal markers and 2 competing events from the primary biliary cholangitis data.}
\label{FigS2}
\end{figure}

\newpage \ \newpage

\subsection*{Appendix A: Details about INLA}
It is out of the scope of this paper to provide all the details of the integrated nested Laplace approximation routine. We propose here an overview and point the readers interested in more details to relevant references. By consulting these resources, one can gain a more complete understanding of INLA and of how it can be used for joint modeling.

INLA is a method for approximating posterior distributions for LGMs, which is a type of statistical model that assumes a multivariate Gaussian distribution for the latent field (i.e., vector of all unknown parameters in the linear predictor). The precision matrix of the latent field depends on the hyperparameters of the model. This method works when this multivariate Gaussian distribution has a sparse precision matrix. This is the case when each element of the latent field is independent of many of the others when conditioning on a few of them, which defines a Gaussian Markov Random Field - GMRF, see \citet{rue2005gaussian} for details. INLA takes advantage of the sparse structure of the precision matrix in LGMs to perform efficient Bayesian inference by considering the state of the art on numerical algorithms for sparse matrices.

The original INLA paper (\cite{Rue09}) describes the INLA methodology and its ability to take advantage of sparse precision matrices, which allows for efficient computation even for models with large latent parameter spaces. It shows how INLA performs as well as MCMC for models with Gaussian latent effects and likelihood identifiability, in shorter computation time for various scenarios. Initially, the most accurate approach in the INLA methodology was a series of Gaussian approximations (similar to the size of the latent space) which is more costly than one Gaussian approximation. In the modern INLA formulation, the series of Gaussian approximations has been circumvented with a low-rank joint variational Bayes correction to the single Gaussian approximation (\citet{van2021correcting}). Additionally, now the linear predictors can be removed from the latent space resulting in an even more efficient, accurate, scalable, and stable approximation of the marginal posteriors (\citet{NiekerkAvenue}). Various numerical advances are also included in the modern INLA framework like the use of a smart gradient (\citet{fattah2021smart}) for the optimization steps. The low-rank correction is based on a joint implicit change to the posterior means of the parameters by explicitly correcting the solution of the linear system associated with the joint Gaussian approximation to the latent space. While the common use of variational approximations assumes some family for the marginal posteriors of the parameters, INLA does not. In fact, a Gaussian family is assumed only for the joint conditional posterior (which is Gaussian-like because we only work with latent Gaussian models) and then we integrate over the hyperparameter space to get the marginals for the latent parameters (which are then not Gaussian). The low-rank correction can be applied in a lower dimension than that of the precision matrix because correcting only some of the conditionals has an impact on all the marginal posteriors. The correction is thus less cumbersome compared to the common variational Bayes that does an approximation to all the marginals. Specifically, the objective of this joint correction is to find the distribution that minimizes the Kullback-Leibler divergence between the true joint posterior distribution and the approximating joint distribution. The modern formulation forms the basis for future developments in INLA.

Another area where INLA has evolved a lot is in the parallel computations, novel parallelization strategies for the INLA methodology using OpenMP have been introduced (\citet{gaedke2022parallelized}). The approach introduces two layers of parallelism, including the parallelization of independent function evaluations and the use of the PARDISO sparse solver, which internally employs OpenMP for simultaneous operations. These advancements allow for the efficient execution of more complex models at higher resolution in shorter runtimes with INLA.

Finally, it is important to detail how joint models fit with the INLA method. In order to do so, one has to show that joint models are latent Gaussian models. It has been shown that any joint model with a linear association structure is a latent Gaussian model, and can therefore be analyzed using INLA (\citet{martino2011approximate, van2019joint}). A joint model consists of two types of submodels: survival or longitudinal. The survival model has a likelihood defined as the product of the likelihoods for each individual, and the longitudinal model has a likelihood defined as the product of the likelihoods for each individual and measurement occasion. Both submodels are linked by random effects which are part of the latent Gaussian field. The hyperparameters are assigned a prior distribution, and the marginal posterior distributions for the joint model are defined as the product of the likelihoods for the survival and longitudinal submodels, the latent field and the prior distribution for the hyperparameters. This construction underlines that the joint model is an LGM, and can therefore be analyzed using tools such as INLA. INLA is particularly useful for analyzing joint models because it takes advantage of the relationship between the outcomes, the approximation error is reduced compared to univariate cross-sectional models because there are multiple data points related to each parameter.

\subsection*{Appendix B: Illustrative example of relative versus absolute error}
In the context of a Poisson GLM with only one parameter which is the rate parameter of the Poisson distribution, the relative error of INLA is inverse proportional to $\mathcal{O}(n^{-3/2})$, where $n$ is the number of data points for a given parameter in the model (see Section 4 of \citet{Rue09} for details and properties for more complex models). We can illustrate this by sampling counts from a Poisson distribution. Figure S3 displays posterior marginals for this rate parameter with MCMC and INLA. The true posterior distribution is obtained by running MCMC for an infinite number of iterations (approximated by running for 1000000 iterations, displayed with the dashed curve in Figure S3). INLA performs poorly when there are only a few data points ($n=5$) and a low rate ($\lambda=1$), which is an extreme case (observed data is 0,0,0,0,1) but performs well with more data ($n=100$) or when the rate is higher ($\lambda=5$) while the approximation error with MCMC depends on the number of iterations, regardless of the number of data points or the rate parameter. Note that in this example, both INLA and MCMC assume the same prior.

\begin{figure}[!ht]
\centering
\includegraphics[scale=0.75]{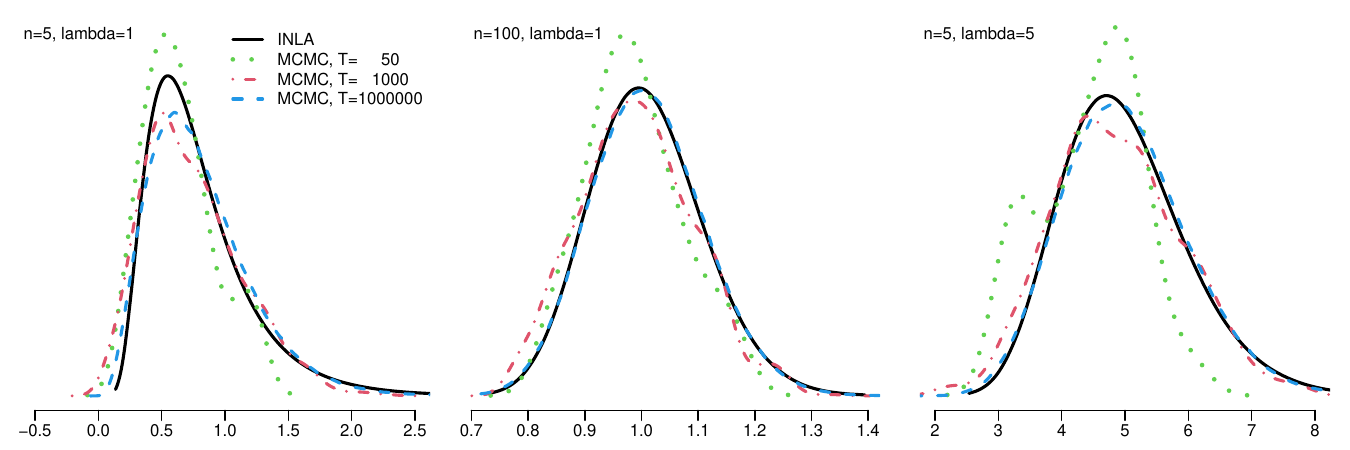}
\caption*{Fig. S3: Posterior marginals for the rate parameter of a Poisson distribution with INLA and MCMC assuming different number of iterations.}
\end{figure}

\subsection*{Appendix C: Bayesian modeling diagnostics}
It is important to do prior sensitivity analysis when fitting Bayesian models. It consists in the evaluation of the impact of priors on the posteriors by fitting the same model with different priors. It is possible to plot priors and posteriors distributions with \textbf{INLAjoint} by adding the argument ``priors=TRUE'' to the \textit{plot()} function call. We illustrate this procedure here by fitting an identifiable model with two different set of priors and by refitting the model with reduced data such that there are identifiability issues. The model is a linear mixed effects model (with fixed and random intercept and slope as well as a binary treatment covariate). We plotted the prior versus posterior for the correlation between the random intercept and the random slope in Figure S4. When the model is identifiable, the prior has a low impact on the posteriors while the model with identifiability issues is very sentitive to prior specification. The code to replicate this prior sensitivity analysis is available on Github (https://github.com/DenisRustand/MultivJoint/blob/main/Priorsensitivity.R)
\begin{figure}[!ht]
\centering
\includegraphics[scale=0.75]{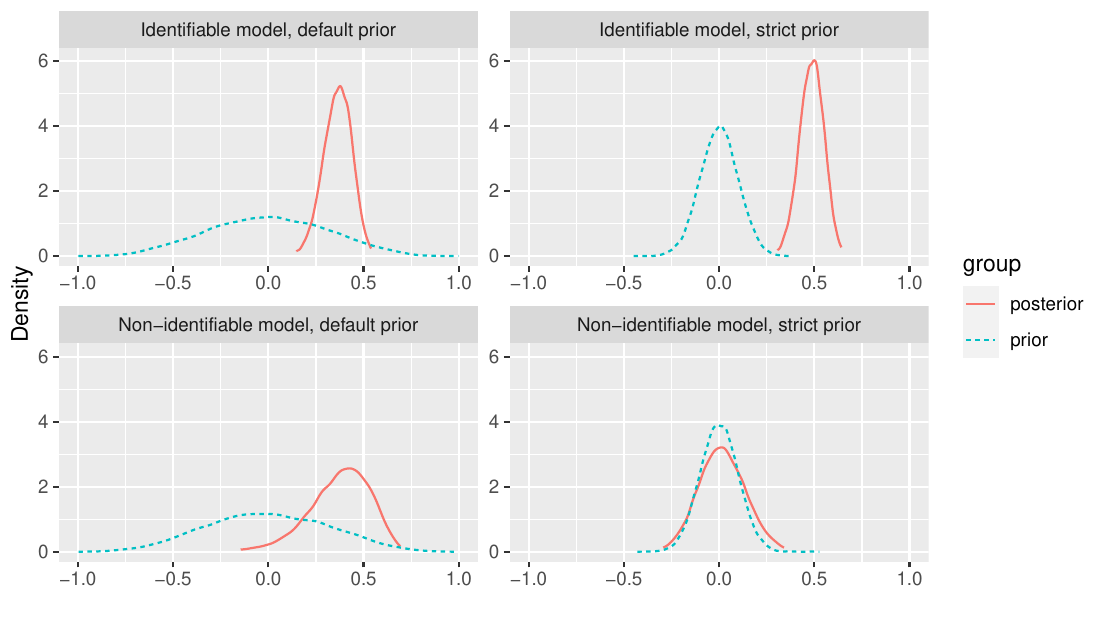}
\caption*{Fig. S4: Comparison of posterior distributions for a correlation parameter with identifiable / non-identifiable models assuming different priors.}
\end{figure}

\newpage

\begin{table}[!p]
\caption*{Table S1: Simulations with $K$=1 continuous longitudinal marker.}
\footnotesize
\centering
{\tabcolsep=1pt
\begin{tabular}{@{}lccccccccccccccccccccccccccccc@{}}
\hline
Approach: &  \multicolumn{3}{c}{R-INLA 1} & \multicolumn{3}{c}{R-INLA 2} & \multicolumn{3}{c}{joineRML*} & \multicolumn{3}{c}{JMbayes2} & \multicolumn{3}{c}{rstanarm 1} & \multicolumn{3}{c}{rstanarm 2}\\
 & \multicolumn{3}{c}{(Empirical Bayes)} & \multicolumn{3}{c}{(Full Bayesian)} & & & & & & & \multicolumn{3}{c}{(1 chain / 1000 iter.)} & \multicolumn{3}{c}{(4 chains / 2000 iter.)}\\
True value & Bias & (SD) & CP (\%) & Bias & (SD) & CP & Bias & (SD) & CP & Bias & (SD) & CP & Bias & (SD) & CP & Bias & (SD) & CP  \\
\hline
$\beta_{10}$=0.2 & 0.002 & (0.054) & 97 & 0.002 & (0.054) & 97 & -0.001 & (0.053) & 96 & 0.003 & (0.054) & 97 & 0.003 & (0.054) & 96 & 0.003 & (0.055) & 96 \\
 $\beta_{11}$=-0.1 & 0 & (0.026) & 95 & 0 & (0.026) & 95 & -0.006 & (0.024) & 92 & 0 & (0.026) & 94 & 0 & (0.026) & 94 & 0 & (0.021) & 96 \\
 $\beta_{12}$=0.1 & -0.002 & (0.045) & 96 & -0.002 & (0.045) & 96 & -0.002 & (0.045) & 95 & -0.003 & (0.045) & 95 & -0.002 & (0.045) & 96 & -0.003 & (0.045) & 95 \\
 $\beta_{13}$=-0.2 & -0.001 & (0.041) & 98 & -0.001 & (0.041) & 98 & -0.001 & (0.041) & 97 & 0 & (0.041) & 97 & -0.001 & (0.041) & 97 & -0.001 & (0.042) & 97 \\
 $\sigma_{\varepsilon 1}$=0.4 & -0.001 & (0.006) & 94 & -0.001 & (0.006) & 94 & -0.001 & (0.006) & 88 & 0 & (0.006) & 96 & 0 & (0.006) & 96 & 0.001 & (0.006) & 92 \\
 $\sigma_{b10}^2$=0.16 & 0.004 & (0.015) & 94 & 0.005 & (0.015) & 94 & -0.007 & (0.016) & 88 & 0 & (0.016) & 97 & 0.001 & (0.016) & 93 & 0.001 & (0.016) & 94 \\
 $\sigma_{b11}^2$=0.16 & 0.006 & (0.017) & 92 & 0.006 & (0.017) & 93 & -0.008 & (0.017) & 91 & 0.001 & (0.018) & 92 & 0.004 & (0.018) & 93 & 0.001 & (0.011) & 97 \\
 $\textrm{cov}_{b10,b11}$=0.08 & -0.004 & (0.012) & 95 & -0.004 & (0.012) & 95 & 0.005 & (0.013) & 90 & -0.003 & (0.012) & 96 & 0 & (0.012) & 95 & 0.001 & (0.01) & 95 \\
 $\varphi_1$=0.5 & 0.006 & (0.081) & 93 & 0.005 & (0.081) & 93 & 0.005 & (0.083) & 94 & 0.01 & (0.081) & 98 & 0.01 & (0.082) & 94 & 0.01 & (0.094) & 94 \\
 Conv. rate & \multicolumn{3}{c}{1} & \multicolumn{3}{c}{1} & \multicolumn{3}{c}{1} & \multicolumn{3}{c}{1} & \multicolumn{3}{c}{1} & \multicolumn{3}{c}{0.98} \\
 Comp. time (sec.) & \multicolumn{3}{c}{5.52 (1.18)} & \multicolumn{3}{c}{5.65 (2.19)} & \multicolumn{3}{c}{19.09 (11.38)} & \multicolumn{3}{c}{47.09 (0.97)} & \multicolumn{3}{c}{271.28 (95.11)} & \multicolumn{3}{c}{505.31 (266.75)}\\
  \arrayrulecolor{black}\hline
 \multicolumn{19}{l}{*In joineRML, the two covariates were further included in the survival model to make the association structure comparable across software}
\end{tabular}}
\label{sim1Y}
\end{table}

\vspace{2cm}

\begin{table}[!p]
\caption*{Table S2: Simulations with $K$=2 continuous longitudinal markers.}
\footnotesize
\centering
{\tabcolsep=1pt
\begin{tabular}{@{}lccccccccccccccccccccccccccccc@{}}
\hline
Approach: &  \multicolumn{3}{c}{R-INLA 1} & \multicolumn{3}{c}{R-INLA 2} & \multicolumn{3}{c}{joineRML*} & \multicolumn{3}{c}{JMbayes2} & \multicolumn{3}{c}{rstanarm 1} & \multicolumn{3}{c}{rstanarm 2}\\
 & \multicolumn{3}{c}{(Empirical Bayes)} & \multicolumn{3}{c}{(Full Bayesian)} & & & & & & & \multicolumn{3}{c}{(1 chain / 1000 iter.)} & \multicolumn{3}{c}{(4 chains / 2000 iter.)}\\
True value & Bias & (SD) & CP (\%) & Bias & (SD) & CP & Bias & (SD) & CP & Bias & (SD) & CP & Bias & (SD) & CP & Bias & (SD) & CP  \\
\hline
$\beta_{10}$=0.2 & -0.001 & (0.051) & 97 & -0.001 & (0.051) & 97 & -0.001 & (0.051) & 96 & 0 & (0.051) & 96 & -0.001 & (0.051) & 96 & -0.002 & (0.051) & 96 \\
 $\beta_{11}$=-0.1 & 0 & (0.026) & 95 & 0 & (0.026) & 96 & -0.011 & (0.026) & 93 & 0.002 & (0.026) & 95 & 0.001 & (0.026) & 94 & 0 & (0.027) & 95 \\
 $\beta_{12}$=0.1 & 0.003 & (0.042) & 95 & 0.003 & (0.042) & 95 & 0.002 & (0.042) & 96 & 0.001 & (0.043) & 95 & 0.003 & (0.042) & 95 & 0.003 & (0.043) & 95 \\
 $\beta_{13}$=-0.2 & -0.001 & (0.04) & 96 & -0.001 & (0.04) & 96 & -0.001 & (0.041) & 97 & -0.001 & (0.04) & 96 & -0.001 & (0.041) & 95 & 0 & (0.04) & 97 \\
 $\sigma_{\varepsilon 1}$=0.4 & -0.001 & (0.006) & 93 & -0.001 & (0.006) & 93 & -0.001 & (0.006) & 89 & 0 & (0.006) & 94 & 0 & (0.007) & 93 & 0 & (0.006) & 93 \\
 $\beta_{20}$=0.2 & 0.002 & (0.065) & 94 & 0.002 & (0.065) & 94 & 0.003 & (0.065) & 95 & 0.002 & (0.065) & 94 & 0.002 & (0.065) & 92 & 0.001 & (0.066) & 93 \\
 $\beta_{21}$=-0.1 & 0.003 & (0.032) & 94 & 0.003 & (0.032) & 95 & 0.019 & (0.032) & 89 & 0.003 & (0.032) & 94 & 0.002 & (0.032) & 93 & 0.004 & (0.033) & 95 \\
 $\beta_{22}$=0.1 & 0.003 & (0.053) & 94 & 0.003 & (0.053) & 94 & 0.003 & (0.054) & 95 & 0.003 & (0.053) & 94 & 0.003 & (0.054) & 94 & 0.004 & (0.054) & 94 \\
 $\beta_{23}$=-0.2 & -0.004 & (0.053) & 92 & -0.005 & (0.053) & 92 & -0.005 & (0.054) & 93 & -0.004 & (0.053) & 92 & -0.005 & (0.053) & 93 & -0.004 & (0.053) & 92 \\
 $\sigma_{\varepsilon 2}$=0.4 & 0 & (0.006) & 95 & 0 & (0.006) & 95 & -0.001 & (0.006) & 89 & 0 & (0.006) & 95 & 0 & (0.006) & 94 & 0 & (0.006) & 94 \\
 $\sigma_{b10}^2$=0.16 & 0.006 & (0.014) & 94 & 0.006 & (0.014) & 94 & -0.004 & (0.015) & 94 & 0.001 & (0.015) & 98 & 0.001 & (0.015) & 94 & 0.001 & (0.015) & 95 \\
 $\sigma_{b11}^2$=0.16 & 0.006 & (0.017) & 96 & 0.007 & (0.018) & 96 & -0.007 & (0.017) & 91 & 0.003 & (0.017) & 95 & 0.003 & (0.019) & 94 & 0.003 & (0.018) & 96 \\
 $\sigma_{b20}^2$=0.25 & 0.004 & (0.021) & 94 & 0.004 & (0.021) & 94 & -0.003 & (0.021) & 93 & 0.003 & (0.022) & 94 & 0.003 & (0.022) & 94 & 0.005 & (0.021) & 94 \\
 $\sigma_{b21}^2$=0.25 & 0.004 & (0.025) & 94 & 0.004 & (0.025) & 95 & -0.008 & (0.025) & 92 & 0.001 & (0.025) & 95 & 0.004 & (0.026) & 95 & 0.003 & (0.025) & 94 \\
 $\textrm{cov}_{b10,b11}$=0.08 & -0.006 & (0.012) & 93 & -0.006 & (0.012) & 93 & 0 & (0.012) & 95 & -0.004 & (0.012) & 93 & -0.002 & (0.013) & 95 & -0.002 & (0.012) & 95 \\
 $\textrm{cov}_{b10,b20}$=0.02 & 0 & (0.012) & 97 & 0 & (0.013) & 97 & -0.001 & (0.012) & 97 & -0.001 & (0.012) & 97 & -0.001 & (0.012) & 96 & -0.001 & (0.012) & 96 \\
 $\textrm{cov}_{b10,b21}$=0.04 & -0.002 & (0.014) & 97 & -0.002 & (0.014) & 98 & -0.002 & (0.014) & 97 & -0.002 & (0.013) & 95 & -0.001 & (0.014) & 95 & -0.001 & (0.014) & 97 \\
 $\textrm{cov}_{b11,b20}$=0.04 & -0.002 & (0.015) & 96 & -0.001 & (0.015) & 96 & -0.001 & (0.015) & 94 & -0.002 & (0.015) & 94 & -0.001 & (0.015) & 95 & -0.001 & (0.015) & 96 \\
 $\textrm{cov}_{b11,b21}$=0 & 0 & (0.016) & 95 & 0 & (0.016) & 95 & -0.001 & (0.015) & 95 & -0.002 & (0.015) & 94 & -0.002 & (0.016) & 90 & -0.002 & (0.016) & 94 \\
 $\textrm{cov}_{b20,b21}$=0.1 & -0.004 & (0.018) & 95 & -0.003 & (0.017) & 95 & -0.001 & (0.018) & 96 & -0.004 & (0.017) & 94 & -0.002 & (0.017) & 96 & -0.002 & (0.017) & 94 \\
 $\varphi_1$=0.5 & -0.004 & (0.087) & 93 & -0.005 & (0.086) & 92 & -0.007 & (0.087) & 95 & -0.005 & (0.088) & 98 & 0.002 & (0.089) & 93 & 0 & (0.088) & 94 \\
 $\varphi_2$=-0.5 & -0.002 & (0.068) & 95 & -0.002 & (0.069) & 94 & 0 & (0.068) & 96 & -0.005 & (0.069) & 96 & -0.009 & (0.07) & 93 & -0.007 & (0.068) & 96 \\
 Conv. rate & \multicolumn{3}{c}{1} & \multicolumn{3}{c}{1} & \multicolumn{3}{c}{1} & \multicolumn{3}{c}{1} & \multicolumn{3}{c}{0.99} & \multicolumn{3}{c}{0.88} \\
 Comp. time (sec.) & \multicolumn{3}{c}{15.26 (6.17)} & \multicolumn{3}{c}{24.89 (10.12)} & \multicolumn{3}{c}{32.27 (14.9)} & \multicolumn{3}{c}{80.19 (2.93)} & \multicolumn{3}{c}{526.26 (150.9)} & \multicolumn{3}{c}{1175.07 (290.25)}\\
 \arrayrulecolor{black}\hline
 \multicolumn{19}{l}{*In joineRML, the two covariates were further included in the survival model to make the association structure comparable across software}
\end{tabular}}
\label{sim2Y}
\end{table}
\newpage

\begin{table}[!p]
\caption*{Table S3: Simulations with $K$=1 count longitudinal marker.}
\footnotesize
\centering
{\tabcolsep=2pt
\begin{tabular}{@{}lccccccccccccccccccccccccccccc@{}}
\hline
Approach: &  \multicolumn{3}{c}{R-INLA 1} & \multicolumn{3}{c}{R-INLA 2} & \multicolumn{3}{c}{JMbayes2} & \multicolumn{3}{c}{rstanarm 1} & \multicolumn{3}{c}{rstanarm 2}\\
 & \multicolumn{3}{c}{(Empirical Bayes)} & \multicolumn{3}{c}{(Full Bayesian)} & & & & \multicolumn{3}{c}{(1 chain / 1000 iter.)} & \multicolumn{3}{c}{(4 chains / 2000 iter.)}\\
True value & Bias & (SD) & CP (\%) & Bias & (SD) & CP & Bias & (SD) & CP & Bias & (SD) & CP & Bias & (SD) & CP \\
\hline
$\beta_{10}$=4 & -0.002 & (0.044) & 95 & -0.002 & (0.044) & 95 & 0 & (0.044) & 95 & 0.008 & (0.056) & 87 & 0.003 & (0.046) & 95 \\
 $\beta_{11}$=-0.1 & 0 & (0.016) & 95 & 0 & (0.016) & 95 & 0.001 & (0.016) & 96 & -0.011 & (0.042) & 91 & -0.006 & (0.025) & 93 \\
 $\beta_{12}$=0.1 & 0.001 & (0.035) & 96 & 0.001 & (0.035) & 96 & 0 & (0.035) & 96 & 0 & (0.037) & 87 & 0.001 & (0.036) & 93 \\
 $\beta_{13}$=-0.2 & 0.002 & (0.034) & 96 & 0.002 & (0.034) & 96 & 0 & (0.034) & 95 & -0.002 & (0.058) & 91 & 0.002 & (0.035) & 95 \\
 $\sigma_{b10}^2$=0.16 & 0.002 & (0.012) & 94 & 0.002 & (0.012) & 94 & 0.002 & (0.012) & 94 & -0.011 & (0.042) & 84 & -0.005 & (0.022) & 94 \\
 $\sigma_{b11}^2$=0.09 & 0.004 & (0.008) & 95 & 0.004 & (0.008) & 95 & 0.001 & (0.008) & 93 & 0.036 & (0.12) & 88 & 0.024 & (0.059) & 95 \\
 $\textrm{cov}_{b10,b11}$=0.06 & -0.001 & (0.008) & 95 & -0.001 & (0.008) & 95 & 0 & (0.007) & 92 & -0.004 & (0.018) & 87 & -0.002 & (0.012) & 92 \\
 $\varphi_1$=0.2 & -0.012 & (0.097) & 93 & -0.015 & (0.093) & 93 & -0.003 & (0.094) & 97 & -0.01 & (0.095) & 93 & -0.009 & (0.093) & 93 \\
 Conv. rate & \multicolumn{3}{c}{1} & \multicolumn{3}{c}{1} & \multicolumn{3}{c}{1} & \multicolumn{3}{c}{0.61} & \multicolumn{3}{c}{0.62} & \\
 Comp. time (sec.) & \multicolumn{3}{c}{7.04 (1.12)} & \multicolumn{3}{c}{7.33 (1.24)} & \multicolumn{3}{c}{92.31 (1.94)} & \multicolumn{3}{c}{724.87 (508.13)} & \multicolumn{3}{c}{2110.49 (1287.46)}\\
  \arrayrulecolor{black}\hline
\end{tabular}}
\label{simp1Y}
\end{table}

\vspace{2cm}

\begin{table}[!p]
\vspace{-1cm}\caption*{Table S4: Simulations with $K=2$ count longitudinal markers.}
\footnotesize
\centering
{\tabcolsep=2pt
\hskip-0.4cm\begin{tabular}{@{}lccccccccccccccccccccccccccccc@{}}
\hline
Approach: &  \multicolumn{3}{c}{R-INLA 1} & \multicolumn{3}{c}{R-INLA 2} & \multicolumn{3}{c}{JMbayes2} & \multicolumn{3}{c}{rstanarm 1} & \multicolumn{3}{c}{rstanarm 2}\\
 & \multicolumn{3}{c}{(Empirical Bayes)} & \multicolumn{3}{c}{(Full Bayesian)} & & & & \multicolumn{3}{c}{(1 chain / 1000 iter.)} & \multicolumn{3}{c}{(4 chains / 2000 iter.)}\\
True value & Bias & (SD) & CP (\%) & Bias & (SD) & CP & Bias & (SD) & CP & Bias & (SD) & CP & Bias & (SD) & CP \\
\hline
$\beta_{10}$=4 & -0.002 & (0.038) & 97 & -0.002 & (0.038) & 97 & 0.001 & (0.038) & 97 & 0.002 & (0.05) & 92 & -0.004 & (0.05) & 96 \\
 $\beta_{11}$=-0.1 & 0 & (0.016) & 96 & 0 & (0.016) & 96 & 0.001 & (0.016) & 96 & -0.007 & (0.038) & 92 & -0.009 & (0.07) & 94 \\
 $\beta_{12}$=0.1 & 0.002 & (0.033) & 96 & 0.002 & (0.033) & 95 & 0.001 & (0.033) & 96 & 0.001 & (0.036) & 89 & 0.006 & (0.045) & 92 \\
 $\beta_{13}$=-0.2 & -0.001 & (0.034) & 95 & -0.001 & (0.034) & 95 & -0.002 & (0.034) & 95 & -0.001 & (0.036) & 89 & 0.001 & (0.036) & 93 \\
 $\beta_{20}$=2 & -0.001 & (0.063) & 94 & -0.001 & (0.063) & 94 & 0.002 & (0.062) & 94 & 0.001 & (0.058) & 93 & -0.003 & (0.071) & 94 \\
 $\beta_{21}$=-0.1 & -0.002 & (0.027) & 95 & -0.001 & (0.027) & 96 & 0 & (0.027) & 96 & -0.003 & (0.029) & 95 & -0.013 & (0.104) & 96 \\
 $\beta_{22}$=0.1 & 0.005 & (0.051) & 95 & 0.005 & (0.051) & 95 & 0.004 & (0.051) & 95 & 0.005 & (0.047) & 90 & 0.013 & (0.099) & 97 \\
 $\beta_{23}$=-0.2 & -0.001 & (0.05) & 95 & -0.001 & (0.05) & 95 & -0.002 & (0.05) & 95 & -0.006 & (0.052) & 92 & -0.01 & (0.116) & 96 \\
 $\sigma_{b10}^2$=0.16 & 0.004 & (0.037) & 96 & 0.002 & (0.011) & 97 & 0.001 & (0.011) & 99 & -0.007 & (0.035) & 90 & -0.003 & (0.019) & 96 \\
 $\sigma_{b11}^2$=0.09 & 0.004 & (0.009) & 94 & 0.004 & (0.008) & 94 & 0.001 & (0.008) & 95 & 0.034 & (0.143) & 93 & 0.016 & (0.056) & 94 \\
 $\sigma_{b20}^2$=0.25 & 0.003 & (0.022) & 94 & 0.004 & (0.022) & 94 & 0.002 & (0.023) & 92 & 0.004 & (0.022) & 95 & 0.004 & (0.025) & 93 \\
 $\sigma_{b21}^2$=0.16 & 0.005 & (0.019) & 92 & 0.005 & (0.019) & 92 & 0.003 & (0.02) & 90 & 0.003 & (0.023) & 92 & 0.004 & (0.021) & 90 \\
 $\textrm{cov}_{b10,b11}$=0.06 & -0.002 & (0.014) & 96 & -0.001 & (0.007) & 96 & -0.001 & (0.007) & 96 & -0.005 & (0.021) & 90 & -0.003 & (0.012) & 93 \\
 $\textrm{cov}_{b10,b20}$=0.02 & 0 & (0.013) & 96 & 0 & (0.011) & 97 & -0.001 & (0.01) & 96 & -0.002 & (0.01) & 89 & -0.003 & (0.012) & 96 \\
 $\textrm{cov}_{b10,b21}$=0.04 & -0.002 & (0.012) & 96 & -0.002 & (0.01) & 96 & -0.001 & (0.01) & 96 & -0.002 & (0.011) & 92 & -0.003 & (0.011) & 97 \\
 $\textrm{cov}_{b11,b20}$=0.03 & -0.002 & (0.011) & 92 & -0.001 & (0.01) & 93 & -0.001 & (0.01) & 92 & 0 & (0.012) & 93 & 0.001 & (0.014) & 95 \\
 $\textrm{cov}_{b11,b21}$=0 & 0 & (0.008) & 97 & 0 & (0.008) & 97 & 0 & (0.008) & 96 & 0 & (0.012) & 95 & 0.002 & (0.012) & 96 \\
 $\textrm{cov}_{b20,b21}$=0.08 & -0.002 & (0.015) & 95 & -0.002 & (0.015) & 95 & 0 & (0.014) & 95 & 0.001 & (0.014) & 96 & 0.001 & (0.016) & 94 \\
 $\varphi_1$=0.2 & -0.002 & (0.091) & 93 & 0 & (0.095) & 93 & 0.008 & (0.091) & 98 & -0.007 & (0.087) & 90 & -0.001 & (0.138) & 95 \\
 $\varphi_2$=-0.2 & -0.004 & (0.073) & 94 & -0.002 & (0.075) & 94 & -0.005 & (0.073) & 96 & 0.007 & (0.065) & 96 & -0.008 & (0.119) & 94 \\
 Conv. rate & \multicolumn{3}{c}{1} & \multicolumn{3}{c}{1} & \multicolumn{3}{c}{1} & \multicolumn{3}{c}{0.25} & \multicolumn{3}{c}{0.44} & \\
 Comp. time (sec.) & \multicolumn{3}{c}{17.19 (2.71)} & \multicolumn{3}{c}{22.75 (2.88)} & \multicolumn{3}{c}{180 (7.82)} & \multicolumn{3}{c}{1905.23 (1042.74)} & \multicolumn{3}{c}{5411.76 (2324.49)}\\
  \arrayrulecolor{black}\hline
\end{tabular}}
\label{simp2Y}
\end{table}

\begin{table}[!p]
\caption*{Table S5: Simulations with $K$=3 count longitudinal markers.}
\footnotesize
\centering
{\tabcolsep=2pt
\hskip-0.4cm\begin{tabular}{@{}lccccccccccccccccccccccccccccc@{}}
\hline
Approach: &  \multicolumn{3}{c}{R-INLA 1} & \multicolumn{3}{c}{R-INLA 2} & \multicolumn{3}{c}{JMbayes2} & \multicolumn{3}{c}{rstanarm 1} & \multicolumn{3}{c}{rstanarm 2}\\
 & \multicolumn{3}{c}{(Empirical Bayes)} & \multicolumn{3}{c}{(Full Bayesian)} & & & & \multicolumn{3}{c}{(1 chain / 1000 iter.)} & \multicolumn{3}{c}{(4 chains / 2000 iter.)}\\
True value & Bias & (SD) & CP (\%) & Bias & (SD) & CP & Bias & (SD) & CP & Bias & (SD) & CP & Bias & (SD) & CP \\
\hline
$\beta_{10}$=4 & -0.002 & (0.043) & 96 & -0.002 & (0.043) & 95 & 0.001 & (0.043) & 95 & -0.021 & (0.469) & 69 & 0.017 & (0.212) & 93 \\
 $\beta_{11}$=-0.1 & -0.001 & (0.017) & 95 & -0.001 & (0.017) & 95 & 0 & (0.017) & 95 & -0.027 & (0.131) & 74 & -0.032 & (0.279) & 95 \\
 $\beta_{12}$=0.1 & -0.001 & (0.035) & 94 & -0.001 & (0.035) & 93 & -0.002 & (0.035) & 93 & 0.013 & (0.147) & 72 & -0.009 & (0.106) & 93 \\
 $\beta_{13}$=-0.2 & 0.004 & (0.032) & 96 & 0.004 & (0.032) & 97 & 0.003 & (0.032) & 96 & -0.013 & (0.154) & 77 & 0.004 & (0.067) & 94 \\
 $\beta_{20}$=2 & -0.006 & (0.064) & 96 & -0.006 & (0.063) & 96 & -0.004 & (0.064) & 95 & 0.013 & (0.229) & 82 & -0.018 & (0.444) & 96 \\
 $\beta_{21}$=-0.1 & 0 & (0.027) & 93 & 0 & (0.028) & 93 & 0.001 & (0.027) & 96 & -0.027 & (0.268) & 81 & -0.008 & (0.134) & 96 \\
 $\beta_{22}$=0.1 & 0.005 & (0.051) & 95 & 0.005 & (0.051) & 95 & 0.004 & (0.051) & 95 & -0.003 & (0.112) & 86 & 0.013 & (0.243) & 97 \\
 $\beta_{23}$=-0.2 & 0.001 & (0.05) & 95 & 0.001 & (0.05) & 95 & 0 & (0.05) & 95 & 0.012 & (0.116) & 85 & 0.002 & (0.052) & 95 \\
 $\beta_{30}$=2 & -0.001 & (0.059) & 95 & -0.001 & (0.059) & 95 & 0.003 & (0.058) & 95 & -0.063 & (0.54) & 85 & 0.002 & (0.067) & 94 \\
 $\beta_{31}$=-0.1 & 0.001 & (0.025) & 94 & 0.001 & (0.026) & 93 & 0.003 & (0.025) & 93 & -0.061 & (0.49) & 85 & -0.003 & (0.035) & 94 \\
 $\beta_{32}$=0.1 & 0.003 & (0.045) & 96 & 0.003 & (0.045) & 96 & 0.001 & (0.045) & 94 & 0.016 & (0.102) & 87 & 0 & (0.055) & 96 \\
 $\beta_{33}$=-0.2 & -0.001 & (0.048) & 95 & -0.001 & (0.048) & 94 & -0.001 & (0.048) & 92 & 0.052 & (0.384) & 85 & 0 & (0.049) & 95 \\
 $\sigma_{b10}^2$=0.16 & 0.002 & (0.012) & 96 & 0.002 & (0.012) & 96 & 0.001 & (0.011) & 99 & -0.035 & (0.062) & 70 & 0.021 & (0.279) & 95 \\
 $\sigma_{b11}^2$=0.09 & 0.004 & (0.009) & 93 & 0.004 & (0.008) & 93 & 0.001 & (0.008) & 95 & 0.088 & (0.187) & 70 & 0.057 & (0.325) & 92 \\
 $\sigma_{b20}^2$=0.25 & 0.002 & (0.021) & 94 & 0.002 & (0.022) & 94 & 0.001 & (0.021) & 93 & 0.053 & (0.641) & 79 & 0.012 & (0.123) & 96 \\
 $\sigma_{b21}^2$=0.16 & 0.004 & (0.018) & 96 & 0.004 & (0.016) & 96 & 0.004 & (0.017) & 95 & 0.026 & (0.143) & 81 & 0.036 & (0.281) & 95 \\
 $\sigma_{b30}^2$=0.25 & 0.008 & (0.022) & 93 & 0.007 & (0.02) & 94 & 0.006 & (0.02) & 93 & 0.016 & (0.168) & 84 & 0.005 & (0.025) & 97 \\
 $\sigma_{b31}^2$=0.16 & 0.007 & (0.016) & 95 & 0.006 & (0.016) & 96 & 0.001 & (0.016) & 96 & 0.013 & (0.084) & 85 & 0.002 & (0.019) & 97 \\
 $\textrm{cov}_{b10,b11}$=0.06 & -0.002 & (0.008) & 95 & -0.002 & (0.007) & 94 & -0.001 & (0.007) & 96 & -0.016 & (0.033) & 68 & -0.005 & (0.028) & 93 \\
 $\textrm{cov}_{b10,b20}$=0.02 & 0.001 & (0.011) & 94 & 0.001 & (0.011) & 94 & 0 & (0.01) & 95 & -0.005 & (0.022) & 70 & -0.001 & (0.012) & 94 \\
 $\textrm{cov}_{b10,b21}$=0.04 & -0.002 & (0.011) & 96 & -0.002 & (0.011) & 96 & -0.003 & (0.011) & 95 & -0.01 & (0.022) & 74 & -0.016 & (0.135) & 96 \\
 $\textrm{cov}_{b10,b30}$=0 & -0.002 & (0.011) & 95 & -0.002 & (0.011) & 96 & 0 & (0.011) & 96 & -0.002 & (0.014) & 75 & -0.001 & (0.012) & 96 \\
 $\textrm{cov}_{b10,b31}$=-0.04 & 0.004 & (0.01) & 96 & 0.004 & (0.009) & 96 & 0.002 & (0.009) & 98 & 0.01 & (0.019) & 77 & 0.003 & (0.011) & 98 \\
 $\textrm{cov}_{b11,b20}$=0.03 & -0.001 & (0.01) & 92 & -0.001 & (0.01) & 92 & -0.002 & (0.01) & 92 & 0.004 & (0.05) & 75 & -0.002 & (0.019) & 89 \\
 $\textrm{cov}_{b11,b21}$=0 & -0.001 & (0.008) & 96 & -0.001 & (0.008) & 97 & -0.003 & (0.008) & 93 & 0.005 & (0.033) & 81 & -0.01 & (0.1) & 96 \\
 $\textrm{cov}_{b11,b30}$=-0.06 & 0.004 & (0.01) & 94 & 0.004 & (0.01) & 94 & -0.001 & (0.01) & 95 & 0.006 & (0.03) & 84 & 0 & (0.016) & 95 \\
 $\textrm{cov}_{b11,b31}$=0 & -0.003 & (0.008) & 96 & -0.003 & (0.007) & 96 & 0.001 & (0.007) & 95 & 0 & (0.037) & 86 & -0.001 & (0.01) & 99 \\
 $\textrm{cov}_{b20,b21}$=0.08 & -0.001 & (0.014) & 95 & -0.002 & (0.014) & 95 & -0.001 & (0.014) & 95 & -0.001 & (0.03) & 82 & 0.001 & (0.016) & 95 \\
 $\textrm{cov}_{b20,b30}$=0.05 & 0.001 & (0.015) & 95 & 0.001 & (0.015) & 96 & -0.002 & (0.014) & 94 & -0.007 & (0.024) & 83 & 0 & (0.017) & 97 \\
 $\textrm{cov}_{b20,b31}$=0.04 & -0.004 & (0.015) & 92 & -0.004 & (0.014) & 92 & -0.005 & (0.014) & 91 & -0.007 & (0.023) & 78 & -0.001 & (0.014) & 93 \\
 $\textrm{cov}_{b21,b30}$=0.04 & -0.001 & (0.017) & 95 & -0.001 & (0.014) & 94 & 0.001 & (0.014) & 94 & 0.003 & (0.024) & 79 & 0.01 & (0.017) & 88 \\
 $\textrm{cov}_{b21,b31}$=-0.04 & 0.003 & (0.012) & 96 & 0.003 & (0.012) & 95 & 0.001 & (0.011) & 96 & 0.005 & (0.021) & 81 & 0.002 & (0.013) & 97 \\
 $\textrm{cov}_{b30,b31}$=0.12 & -0.009 & (0.017) & 90 & -0.008 & (0.015) & 91 & -0.003 & (0.015) & 93 & -0.007 & (0.043) & 86 & -0.001 & (0.017) & 94 \\
 $\varphi_1$=0.2 & -0.007 & (0.091) & 92 & -0.008 & (0.091) & 92 & 0.003 & (0.093) & 98 & -0.078 & (0.471) & 86 & -0.027 & (0.121) & 93 \\
 $\varphi_2$=-0.2 & 0.003 & (0.074) & 95 & 0.002 & (0.074) & 95 & 0.005 & (0.073) & 96 & 0.057 & (0.415) & 90 & -0.007 & (0.08) & 96 \\
 $\varphi_3$=0.2 & 0 & (0.07) & 93 & 0 & (0.07) & 93 & 0.002 & (0.071) & 93 & -0.041 & (0.425) & 90 & 0.008 & (0.132) & 96 \\
 Conv. rate & \multicolumn{3}{c}{1} & \multicolumn{3}{c}{1} & \multicolumn{3}{c}{1} & \multicolumn{3}{c}{0.61} & \multicolumn{3}{c}{0.41} & \\
 Comp. time (sec.) & \multicolumn{3}{c}{46.63 (7.09)} & \multicolumn{3}{c}{81.72 (7.7)} & \multicolumn{3}{c}{319.61 (16.75)} & \multicolumn{3}{c}{3327.91 (977.34)} & \multicolumn{3}{c}{8494.53 (2925.53)}\\
  \arrayrulecolor{black}\hline
\end{tabular}}
\label{simp3Y}
\end{table}

\newpage
\begin{table}[!p]
\vspace{-1cm}\caption*{Table S6: Simulations with $K$=1 binary longitudinal marker.}
\footnotesize
\centering
{\tabcolsep=2pt
\begin{tabular}{@{}lccccccccccccccccccccccccccccc@{}}
\hline
Approach: &  \multicolumn{3}{c}{R-INLA 1} & \multicolumn{3}{c}{R-INLA 2} & \multicolumn{3}{c}{rstanarm 1} & \multicolumn{3}{c}{rstanarm 2}\\
 & \multicolumn{3}{c}{(Empirical Bayes)} & \multicolumn{3}{c}{(Full Bayesian)} &\multicolumn{3}{c}{(1 chain / 1000 iter.)} & \multicolumn{3}{c}{(4 chains / 2000 iter.)}\\
True value & Bias & (SD) & CP (\%) & Bias & (SD) & CP & Bias & (SD) & CP & Bias & (SD) & CP  \\
\hline
$\beta_{10}$=1 & 0.002 & (0.094) & 96 & 0.003 & (0.094) & 96 & 0.014 & (0.095) & 94 & 0.014 & (0.095) & 94 \\
 $\beta_{11}$=-1 & 0.007 & (0.033) & 91 & 0.007 & (0.033) & 91 & -0.003 & (0.034) & 92 & -0.003 & (0.034) & 94 \\
 $\beta_{12}$=1 & -0.017 & (0.078) & 93 & -0.019 & (0.078) & 94 & -0.008 & (0.079) & 94 & -0.008 & (0.079) & 93 \\
 $\beta_{13}$=-1 & 0.009 & (0.075) & 94 & 0.01 & (0.075) & 94 & -0.002 & (0.076) & 95 & -0.001 & (0.076) & 95 \\
 $\sigma_{b10}^2$=0.25 & -0.017 & (0.037) & 92 & -0.017 & (0.037) & 92 & 0.007 & (0.043) & 91 & 0.005 & (0.043) & 92 \\
 $\varphi_1$=0.3 & 0.009 & (0.086) & 95 & 0.01 & (0.086) & 95 & 0.005 & (0.086) & 96 & 0.005 & (0.085) & 96 \\
 Conv. rate & \multicolumn{3}{c}{1} & \multicolumn{3}{c}{1} & \multicolumn{3}{c}{1} & \multicolumn{3}{c}{1} & \\
 Comp. time (sec.) & \multicolumn{3}{c}{4.76 (0.35)} & \multicolumn{3}{c}{5.12 (0.37)} & \multicolumn{3}{c}{155.03 (8.95)} & \multicolumn{3}{c}{235.32 (20.43)}\\
  \arrayrulecolor{black}\hline
\end{tabular}}
\label{simb1Y}
\end{table}

\begin{table}[!p]
\caption*{Table S7: Simulations with $K$=2 binary longitudinal markers.}
\footnotesize
\centering
{\tabcolsep=2pt
\begin{tabular}{@{}lccccccccccccccccccccccccccccc@{}}
\hline
Approach: &  \multicolumn{3}{c}{R-INLA 1} & \multicolumn{3}{c}{R-INLA 2} & \multicolumn{3}{c}{JMbayes2} &\multicolumn{3}{c}{rstanarm 1} & \multicolumn{3}{c}{rstanarm 2}\\
 & \multicolumn{3}{c}{(Empirical Bayes)} & \multicolumn{3}{c}{(Full Bayesian)} & & & & \multicolumn{3}{c}{(1 chain / 1000 iter.)} & \multicolumn{3}{c}{(4 chains / 2000 iter.)}\\
True value & Bias & (SD) & CP (\%) & Bias & (SD) & CP & Bias & (SD) & CP & Bias & (SD) & CP & Bias & (SD) & CP  \\
\hline
$\beta_{10}$=1 & -0.01 & (0.087) & 96 & -0.01 & (0.087) & 98 & -0.001 & (0.088) & 98 & -0.001 & (0.087) & 97 & -0.001 & (0.087) & 97 \\
 $\beta_{11}$=-1 & 0.008 & (0.033) & 93 & 0.008 & (0.033) & 93 & -0.003 & (0.033) & 94 & 0 & (0.034) & 95 & 0 & (0.034) & 96 \\
 $\beta_{12}$=1 & -0.009 & (0.073) & 95 & -0.01 & (0.073) & 95 & -0.004 & (0.074) & 95 & -0.002 & (0.073) & 96 & -0.002 & (0.073) & 96 \\
 $\beta_{13}$=-1 & 0.016 & (0.07) & 95 & 0.017 & (0.07) & 95 & 0.008 & (0.07) & 96 & 0.007 & (0.07) & 95 & 0.008 & (0.07) & 96 \\
 $\beta_{20}$=1 & -0.008 & (0.097) & 93 & -0.008 & (0.096) & 94 & -0.001 & (0.097) & 94 & 0 & (0.097) & 94 & 0 & (0.096) & 94 \\
 $\beta_{21}$=-1 & 0.008 & (0.031) & 92 & 0.009 & (0.031) & 93 & 0.004 & (0.031) & 95 & 0.001 & (0.031) & 96 & 0.001 & (0.031) & 96 \\
 $\beta_{22}$=1 & 0 & (0.081) & 93 & -0.002 & (0.08) & 93 & 0.007 & (0.081) & 93 & 0.007 & (0.081) & 93 & 0.006 & (0.081) & 94 \\
 $\beta_{23}$=-1 & 0 & (0.072) & 95 & 0.001 & (0.071) & 95 & -0.007 & (0.072) & 96 & -0.008 & (0.072) & 95 & -0.008 & (0.072) & 96 \\
 $\sigma_{b10}^2$=0.25 & -0.005 & (0.035) & 97 & -0.005 & (0.035) & 97 & -0.003 & (0.039) & 97 & 0.004 & (0.039) & 96 & 0.003 & (0.039) & 96 \\
 $\sigma_{b20}^2$=0.25 & -0.005 & (0.033) & 97 & -0.005 & (0.033) & 96 & -0.004 & (0.038) & 93 & 0.003 & (0.038) & 96 & 0.002 & (0.038) & 96 \\
 $\textrm{cov}_{b10,b20}$=0.15 & -0.015 & (0.025) & 94 & -0.014 & (0.025) & 95 & -0.011 & (0.026) & 95 & 0 & (0.028) & 95 & 0 & (0.028) & 95 \\
 $\varphi_1$=0.3 & -0.013 & (0.351) & 91 & -0.02 & (0.364) & 91 & -0.147 & (0.319) & 90 & 0.041 & (0.417) & 95 & 0.041 & (0.412) & 95 \\
 $\varphi_2$=-0.3 & 0.02 & (0.351) & 90 & 0.027 & (0.365) & 90 & 0.161 & (0.315) & 90 & -0.037 & (0.415) & 95 & -0.037 & (0.411) & 95 \\
 Conv. rate & \multicolumn{3}{c}{1} & \multicolumn{3}{c}{1} & \multicolumn{3}{c}{1} & \multicolumn{3}{c}{1} & \multicolumn{3}{c}{1} & \\
 Comp. time (sec.) & \multicolumn{3}{c}{10.99 (4.16)} & \multicolumn{3}{c}{11.86 (2.03)} & \multicolumn{3}{c}{89.43 (3.08)} & \multicolumn{3}{c}{405.54 (90.36)} & \multicolumn{3}{c}{825.11 (166.98)}\\
  \arrayrulecolor{black}\hline
\end{tabular}}
\label{simb2Y}
\end{table}

\begin{table}[!p]
\caption*{Table S8: Simulations with $K$=3 binary longitudinal markers.}
\footnotesize
\centering
{\tabcolsep=2pt
\begin{tabular}{@{}lccccccccccccccccccccccccccccc@{}}
\hline
Approach: &  \multicolumn{3}{c}{R-INLA 1} & \multicolumn{3}{c}{R-INLA 2} & \multicolumn{3}{c}{JMbayes2} &\multicolumn{3}{c}{rstanarm 1} & \multicolumn{3}{c}{rstanarm 2}\\
 & \multicolumn{3}{c}{(Empirical Bayes)} & \multicolumn{3}{c}{(Full Bayesian)} & & & & \multicolumn{3}{c}{(1 chain / 1000 iter.)} & \multicolumn{3}{c}{(4 chains / 2000 iter.)}\\
True value & Bias & (SD) & CP (\%) & Bias & (SD) & CP & Bias & (SD) & CP & Bias & (SD) & CP & Bias & (SD) & CP  \\
\hline
$\beta_{10}$=1 & -0.008 & (0.085) & 95 & -0.008 & (0.084) & 96 & -0.001 & (0.084) & 95 & -0.001 & (0.084) & 96 & -0.002 & (0.083) & 96 \\
 $\beta_{11}$=-1 & 0.007 & (0.031) & 92 & 0.007 & (0.031) & 92 & 0.004 & (0.032) & 94 & 0.001 & (0.032) & 95 & 0.001 & (0.032) & 95 \\
 $\beta_{12}$=1 & -0.007 & (0.064) & 95 & -0.008 & (0.064) & 96 & -0.002 & (0.064) & 95 & -0.002 & (0.064) & 96 & -0.001 & (0.065) & 96 \\
 $\beta_{13}$=-1 & 0.014 & (0.069) & 93 & 0.015 & (0.068) & 93 & 0.006 & (0.069) & 94 & 0.008 & (0.069) & 93 & 0.009 & (0.068) & 94 \\
 $\beta_{20}$=1 & 0.002 & (0.092) & 94 & 0.001 & (0.091) & 94 & 0.01 & (0.092) & 94 & 0.011 & (0.092) & 95 & 0.01 & (0.093) & 94 \\
 $\beta_{21}$=-1 & 0.006 & (0.029) & 95 & 0.008 & (0.028) & 95 & -0.006 & (0.029) & 96 & -0.003 & (0.029) & 97 & -0.003 & (0.029) & 97 \\
 $\beta_{22}$=1 & -0.011 & (0.074) & 93 & -0.012 & (0.074) & 93 & -0.001 & (0.074) & 95 & -0.004 & (0.075) & 95 & -0.003 & (0.074) & 96 \\
 $\beta_{23}$=-1 & 0.009 & (0.079) & 90 & 0.01 & (0.078) & 90 & -0.001 & (0.079) & 91 & 0 & (0.079) & 91 & -0.001 & (0.08) & 92 \\
 $\beta_{30}$=1 & -0.01 & (0.089) & 97 & -0.011 & (0.089) & 97 & -0.003 & (0.089) & 97 & -0.002 & (0.089) & 97 & -0.002 & (0.09) & 98 \\
 $\beta_{31}$=-1 & 0.007 & (0.029) & 95 & 0.007 & (0.029) & 96 & -0.004 & (0.029) & 96 & 0 & (0.03) & 97 & 0 & (0.03) & 97 \\
 $\beta_{32}$=1 & -0.005 & (0.073) & 95 & -0.006 & (0.072) & 95 & 0.002 & (0.073) & 95 & 0.002 & (0.073) & 95 & 0.003 & (0.073) & 96 \\
 $\beta_{33}$=-1 & 0.013 & (0.073) & 95 & 0.014 & (0.073) & 95 & 0.006 & (0.073) & 96 & 0.004 & (0.074) & 96 & 0.004 & (0.073) & 96 \\
 $\sigma_{b10}^2$=0.16 & 0.007 & (0.025) & 97 & 0.008 & (0.026) & 97 & -0.003 & (0.03) & 97 & -0.001 & (0.031) & 95 & -0.001 & (0.03) & 94 \\
 $\sigma_{b20}^2$=0.25 & -0.008 & (0.037) & 95 & -0.008 & (0.037) & 95 & 0.004 & (0.041) & 93 & 0.003 & (0.041) & 95 & 0.004 & (0.041) & 95 \\
 $\sigma_{b30}^2$=0.25 & 0 & (0.037) & 95 & 0 & (0.037) & 95 & -0.005 & (0.041) & 92 & 0.003 & (0.042) & 93 & 0.003 & (0.042) & 94 \\
 $\textrm{cov}_{b10,b20}$=0.02 & 0 & (0.022) & 96 & 0 & (0.022) & 96 & -0.003 & (0.022) & 96 & 0 & (0.023) & 96 & 0 & (0.023) & 95 \\
 $\textrm{cov}_{b10,b30}$=0.12 & -0.015 & (0.024) & 91 & -0.015 & (0.023) & 91 & -0.013 & (0.024) & 90 & -0.003 & (0.026) & 93 & -0.002 & (0.026) & 94 \\
 $\textrm{cov}_{b20,b30}$=0.05 & -0.002 & (0.025) & 97 & -0.003 & (0.025) & 96 & -0.003 & (0.025) & 94 & 0 & (0.026) & 94 & 0 & (0.026) & 96 \\
 $\varphi_1$=0.3 & 0.007 & (0.4) & 91 & -0.005 & (0.4) & 90 & 0.296 & (0.409) & 94 & 0.03 & (0.517) & 96 & 0.03 & (0.514) & 97 \\
 $\varphi_2$=-0.3 & -0.003 & (0.235) & 92 & 0.003 & (0.242) & 93 & -0.074 & (0.258) & 96 & -0.026 & (0.249) & 96 & -0.025 & (0.25) & 96 \\
 $\varphi_3$=0.3 & 0.008 & (0.385) & 91 & 0.013 & (0.381) & 92 & -0.197 & (0.37) & 97 & 0.01 & (0.489) & 96 & 0.01 & (0.485) & 97 \\
 Conv. rate & \multicolumn{3}{c}{1} & \multicolumn{3}{c}{1} & \multicolumn{3}{c}{1} & \multicolumn{3}{c}{1} & \multicolumn{3}{c}{0.98} & \\
 Comp. time (sec.) & \multicolumn{3}{c}{25.9 (5.49)} & \multicolumn{3}{c}{29.59 (5.32)} & \multicolumn{3}{c}{147.69 (4.63)} & \multicolumn{3}{c}{599.04 (113.33)} & \multicolumn{3}{c}{1492.44 (311.03)}\\
  \arrayrulecolor{black}\hline
\end{tabular}}
\label{simb3Y}
\end{table}

\begin{table}[!p]
\caption*{Table S9: Application of the multivariate joint model fitted with \textbf{R-INLA} to primary biliary cholangitis.}
\footnotesize
\centering
{\tabcolsep=2.25pt
\begin{tabular}{@{}lccccccccccccccccccccccccccccc@{}}
\hline
Parameter & Mean & SD & Cred. Int. & Parameter & Mean & SD & Cred. Int. & Parameter & Mean & SD & Cred. Int.\\
\hline
$\beta_{10}$ & 0.6 & (0.08) & [0.44 ; 0.76] &$\sigma_{b20}^2$ & 0.17 & (0.02) & [0.13 ; 0.22] &$\sigma_{b41}^2$ & 1.32 & (0.22) & [0.96 ; 1.82] \\
 $\beta_{11}$ & -0.11 & (0.11) & [-0.33 ; 0.11] &$\sigma_{b21}^2$ & 0.21 & (0.05) & [0.13 ; 0.31] &$\sigma_{b42}^2$ & 5.01 & (1.09) & [3.25 ; 7.45] \\
 $\beta_{12}$ & 1.12 & (0.15) & [0.83 ; 1.4] &$\sigma_{b22}^2$ & 0.31 & (0.08) & [0.2 ; 0.5] &$\sigma_{b43}^2$ & 16.37 & (3.86) & [10.13 ; 24.94] \\
 $\beta_{13}$ & 1.77 & (0.18) & [1.42 ; 2.11] &$\sigma_{b23}^2$ & 0.22 & (0.07) & [0.11 ; 0.38] &$\textrm{cov}_{b40,b41}$ & -0.04 & (0.04) & [-0.11 ; 0.04] \\
 $\beta_{14}$ & 1.71 & (0.22) & [1.28 ; 2.15] &$\textrm{cov}_{b20,b21}$ & -0.02 & (0.02) & [-0.07 ; 0.03] &$\textrm{cov}_{b40,b42}$ & -0.08 & (0.08) & [-0.24 ; 0.07] \\
 $\beta_{15}$ & 0.11 & (0.21) & [-0.3 ; 0.51] &$\textrm{cov}_{b20,b22}$ & 0.02 & (0.03) & [-0.05 ; 0.08] &$\textrm{cov}_{b40,b43}$ & -0.06 & (0.14) & [-0.35 ; 0.22] \\
 $\beta_{16}$ & -0.3 & (0.25) & [-0.79 ; 0.19] &$\textrm{cov}_{b20,b23}$ & -0.01 & (0.04) & [-0.09 ; 0.07] &$\textrm{cov}_{b41,b42}$ & -1.83 & (0.44) & [-2.82 ; -1.11] \\
 $\beta_{17}$ & -0.27 & (0.31) & [-0.89 ; 0.34] &$\textrm{cov}_{b21,b22}$ & 0.17 & (0.04) & [0.1 ; 0.25] &$\textrm{cov}_{b41,b43}$ & -3.71 & (0.84) & [-5.61 ; -2.35] \\
 $\sigma_{\varepsilon 1}$ & 0.28 & (0) & [0.27 ; 0.29] &$\textrm{cov}_{b21,b23}$ & 0.09 & (0.04) & [0.01 ; 0.17] &$\textrm{cov}_{b42,b43}$ & 8.65 & (2.04) & [5.36 ; 13.22] \\
 $\sigma_{b10}^2$ & 1.04 & (0.15) & [0.83 ; 1.39] &$\textrm{cov}_{b22,b23}$ & 0.15 & (0.07) & [0.05 ; 0.31] &$\beta_{50}$ & -1.39 & (0.24) & [-1.86 ; -0.92] \\
 $\sigma_{b11}^2$ & 1.59 & (0.27) & [1.13 ; 2.18] &$\beta_{30}$ & 3.55 & (0.03) & [3.49 ; 3.61] &$\beta_{51}$ & -0.1 & (0.34) & [-0.77 ; 0.56] \\
 $\sigma_{b12}^2$ & 2.44 & (0.53) & [1.68 ; 3.72] &$\beta_{31}$ & 0 & (0.05) & [-0.09 ; 0.09] &$\beta_{52}$ & 0.15 & (0.06) & [0.03 ; 0.26] \\
 $\sigma_{b13}^2$ & 1.79 & (0.54) & [0.97 ; 3.05] &$\beta_{32}$ & -0.1 & (0.01) & [-0.12 ; -0.07] &$\beta_{53}$ & -0.03 & (0.09) & [-0.2 ; 0.14] \\
 $\textrm{cov}_{b10,b11}$ & 0.32 & (0.16) & [0.01 ; 0.64] &$\beta_{33}$ & 0 & (0.02) & [-0.03 ; 0.04] &$\sigma_{b50}^2$ & 5.93 & (1.19) & [4.09 ; 8.68] \\
 $\textrm{cov}_{b10,b12}$ & 0.63 & (0.24) & [0.22 ; 1.2] &$\sigma_{\varepsilon 3}$ & 0.31 & (0.01) & [0.3 ; 0.32] &$\sigma_{b51}^2$ & 0.14 & (0.03) & [0.08 ; 0.21] \\
 $\textrm{cov}_{b10,b13}$ & 0.63 & (0.29) & [0.16 ; 1.32] &$\sigma_{b30}^2$ & 0.13 & (0.01) & [0.1 ; 0.15] &$\textrm{cov}_{b50,b51}$ & -0.25 & (0.16) & [-0.63 ; 0] \\
 $\textrm{cov}_{b11,b12}$ & 1.76 & (0.3) & [1.25 ; 2.45] &$\sigma_{b31}^2$ & 0.01 & (0) & [0.01 ; 0.01] &$\gamma_{1}$ & -0.07 & (0.19) & [-0.45 ; 0.3] \\
 $\textrm{cov}_{b11,b13}$ & 0.89 & (0.32) & [0.29 ; 1.56] &$\textrm{cov}_{b30,b31}$ & 0 & (0) & [-0.01 ; 0] &$\gamma_{2}$ & -0.42 & (0.38) & [-1.17 ; 0.33] \\
 $\textrm{cov}_{b12,b13}$ & 1.42 & (0.51) & [0.65 ; 2.64] &$\beta_{40}$ & 5.52 & (0.03) & [5.46 ; 5.58] &$\varphi_1$ & 1.22 & (0.08) & [1.05 ; 1.37] \\
 $\beta_{20}$ & 4.75 & (0.04) & [4.68 ; 4.83] &$\beta_{41}$ & -0.06 & (0.04) & [-0.14 ; 0.03] &$\varphi_2$ & 0.89 & (0.3) & [0.3 ; 1.53] \\
 $\beta_{21}$ & -0.09 & (0.05) & [-0.19 ; 0.02] &$\beta_{42}$ & -0.19 & (0.12) & [-0.41 ; 0.04] &$\varphi_3$ & 1.15 & (0.1) & [0.96 ; 1.35] \\
 $\beta_{22}$ & -0.13 & (0.08) & [-0.29 ; 0.02] &$\beta_{43}$ & -0.95 & (0.24) & [-1.42 ; -0.47] &$\varphi_4$ & -0.35 & (0.16) & [-0.66 ; -0.01] \\
 $\beta_{23}$ & 0.07 & (0.09) & [-0.11 ; 0.25] &$\beta_{44}$ & -1.27 & (0.45) & [-2.15 ; -0.39] &$\varphi_5$ & -1.82 & (0.18) & [-2.17 ; -1.46] \\
 $\beta_{24}$ & -0.02 & (0.13) & [-0.27 ; 0.22] &$\beta_{45}$ & 0.2 & (0.16) & [-0.12 ; 0.51] &$\varphi_6$ & -1.14 & (0.24) & [-1.61 ; -0.63] \\
 $\beta_{25}$ & 0.1 & (0.11) & [-0.11 ; 0.31] &$\beta_{46}$ & -0.44 & (0.35) & [-1.11 ; 0.24] &$\varphi_7$ & -0.63 & (0.08) & [-0.79 ; -0.47] \\
 $\beta_{26}$ & -0.19 & (0.13) & [-0.45 ; 0.07] &$\beta_{47}$ & -0.55 & (0.64) & [-1.8 ; 0.7] &$\varphi_8$ & -0.27 & (0.16) & [-0.6 ; 0.03] \\
 $\beta_{27}$ & 0.04 & (0.17) & [-0.3 ; 0.39] &$\sigma_{b40}^2$ & 0.15 & (0.01) & [0.12 ; 0.17] &$\varphi_9$ & -0.01 & (0.02) & [-0.05 ; 0.04] \\
 $\sigma_{\varepsilon 2}$ & 0.26 & (0) & [0.26 ; 0.27]\\
\end{tabular}}
\label{resapp}
\end{table}

\begin{table}[!p]
\caption*{Table S10: Application of the multivariate joint model fitted with \textbf{R-INLA} to primary biliary cholangitis with full covariance between the 16 random effects (fixed effects only).}
\footnotesize
\centering
{\tabcolsep=2.25pt
\begin{tabular}{@{}lccccccccccccccccccccccccccccc@{}}
\hline
Parameter & Mean & SD & Cred. Int. & Parameter & Mean & SD & Cred. Int.\\
\hline
$\beta_{10}$ & 0.6 & (0.08) & [0.44 ; 0.75] &$\beta_{40}$ & 5.52 & (0.03) & [5.46 ; 5.58] \\
 $\beta_{11}$ & -0.11 & (0.11) & [-0.33 ; 0.11] &$\beta_{41}$ & -0.06 & (0.04) & [-0.14 ; 0.03] \\
 $\beta_{12}$ & 1.16 & (0.14) & [0.89 ; 1.43] &$\beta_{42}$ & -0.17 & (0.11) & [-0.38 ; 0.04] \\
 $\beta_{13}$ & 1.7 & (0.16) & [1.39 ; 2.02] &$\beta_{43}$ & -1.4 & (0.22) & [-1.82 ; -0.97] \\
 $\beta_{14}$ & 1.58 & (0.17) & [1.24 ; 1.92] &$\beta_{44}$ & -2.08 & (0.4) & [-2.86 ; -1.3] \\
 $\beta_{15}$ & 0.12 & (0.19) & [-0.26 ; 0.49] &$\beta_{45}$ & 0.22 & (0.15) & [-0.08 ; 0.51] \\
 $\beta_{16}$ & -0.04 & (0.23) & [-0.48 ; 0.4] &$\beta_{46}$ & -0.37 & (0.31) & [-0.98 ; 0.23] \\
 $\beta_{17}$ & 0.16 & (0.24) & [-0.32 ; 0.63] &$\beta_{47}$ & -0.46 & (0.57) & [-1.57 ; 0.65] \\
 $\sigma_{\varepsilon 1}$ & 0.28 & (0) & [0.28 ; 0.29] &$\beta_{50}$ & -1.49 & (0.26) & [-2 ; -0.99] \\
 $\beta_{20}$ & 4.76 & (0.04) & [4.69 ; 4.83] &$\beta_{51}$ & -0.07 & (0.36) & [-0.79 ; 0.64] \\
 $\beta_{21}$ & -0.09 & (0.05) & [-0.18 ; 0.01] &$\beta_{52}$ & 0.33 & (0.06) & [0.21 ; 0.45] \\
 $\beta_{22}$ & 0.11 & (0.08) & [-0.04 ; 0.27] &$\beta_{53}$ & -0.09 & (0.09) & [-0.26 ; 0.08] \\
 $\beta_{23}$ & 0.27 & (0.09) & [0.1 ; 0.44] &$\gamma_{1}$ & -0.09 & (0.2) & [-0.48 ; 0.29] \\
 $\beta_{24}$ & 0.11 & (0.12) & [-0.11 ; 0.34] &$\gamma_{2}$ & -0.45 & (0.38) & [-1.2 ; 0.3] \\
 $\beta_{25}$ & 0.02 & (0.11) & [-0.19 ; 0.24] &$\varphi_1$ & 1.09 & (0.04) & [0.99 ; 1.19] \\
 $\beta_{26}$ & -0.36 & (0.12) & [-0.6 ; -0.12] &$\varphi_2$ & 0.86 & (0.2) & [0.45 ; 1.25] \\
 $\beta_{27}$ & -0.15 & (0.16) & [-0.46 ; 0.16] &$\varphi_3$ & 1.19 & (0.05) & [1.08 ; 1.3] \\
 $\sigma_{\varepsilon 2}$ & 0.26 & (0) & [0.26 ; 0.27] &$\varphi_4$ & -0.25 & (0.07) & [-0.43 ; -0.07] \\
 $\beta_{30}$ & 3.55 & (0.03) & [3.49 ; 3.61] &$\varphi_5$ & -2.07 & (0.09) & [-2.26 ; -1.9] \\
 $\beta_{31}$ & 0 & (0.05) & [-0.09 ; 0.09] &$\varphi_6$ & -1.18 & (0.23) & [-1.63 ; -0.72] \\
 $\beta_{32}$ & -0.11 & (0.01) & [-0.14 ; -0.09] &$\varphi_7$ & -0.49 & (0.06) & [-0.62 ; -0.35] \\
 $\beta_{33}$ & 0.01 & (0.02) & [-0.02 ; 0.04] &$\varphi_8$ & -0.35 & (0.1) & [-0.57 ; -0.14] \\
 $\sigma_{\varepsilon 3}$ & 0.31 & (0) & [0.31 ; 0.31] &$\varphi_9$ & 0.04 & (0.01) & [0.01 ; 0.08] \\
\end{tabular}}
\label{resapp2}
\end{table}

\end{document}